\documentclass[aps,prl,twocolumn,longbibliography,superscriptaddress]{revtex4-2}
\usepackage{amssymb}
\usepackage{amsbsy}
\usepackage{amsmath}
\usepackage{graphicx}
\usepackage{graphics}
\usepackage{setspace}
\usepackage{array}
\usepackage{color}
\usepackage{fontenc}
\usepackage{textcomp}
\usepackage{bm}
\usepackage{float}
\usepackage{pifont}
\usepackage[bookmarks=false,linkcolor=blue,urlcolor=blue,colorlinks,citecolor=blue]{hyperref}
\usepackage{soul}
\usepackage[T1]{fontenc}
\usepackage[percent]{overpic}

\begin{document}
\author{T. Farajollahpour}
\email{tohid.f@brocku.ca}
\affiliation{Department of Physics, Brock University, St. Catharines, Ontario L2S 3A1, Canada}

\author{R. Ganesh}
\email{r.ganesh@brocku.ca}
\affiliation{Department of Physics, Brock University, St. Catharines, Ontario L2S 3A1, Canada}

\author{K. V. Samokhin}
\email{ksamokhin@brocku.ca}
\affiliation{Department of Physics, Brock University, St. Catharines, Ontario L2S 3A1, Canada}

\date{\today}
	
\title{Light-induced charge and spin Hall currents in materials with $C_4K$ symmetry}
	
\begin{abstract}
Berry curvature can manifest itself in current responses. It has recently been shown that a Berry-curvature quadrupole induces a third-order ac Hall response in systems that break time reversal ($K$) and a fourfold rotational ($C_4$)
symmetries, while remaining invariant under their combination ($C_4K$). In this work, we
demonstrate that incident light can induce a dc Hall current in such systems. We consider a combination of static and light-induced electric fields. We calculate the current perpendicular to both the static field and the fourfold
axis. Remarkably, the induced current is generically spin-polarized. A net charge current appears for
light that is linearly or elliptically polarized, but not for circular polarization. In contrast, the spin
current remains unchanged when the polarization is varied. This allows for rich possibilities
such as generating a pure spin current using circularly polarized light. We demonstrate this
physics using a two-dimensional toy model with $C_4K$ symmetry.

\end{abstract}

\maketitle

Current responses provide an experimentally accessible window into the Berry curvature properties of electronic bands~\cite{Resta_2000,BerryRev}. In particular, the quantum Hall effect and the anomalous Hall effect serve as powerful tools across material families~\cite{Anoma1982,Anoma1985,Anoma1988,AnomalRev2010,MacDonald2014,Nayak2016,AnomaExp1,AnomaExp2}. The latter has been conceptualized as an expansion in moments of the Berry curvature distribution over the Brillouin zone -- the $n^\mathrm{th}$ moment yields a response at the $(n+1)^\mathrm{th}$ order in the applied electric field~\cite{Du2021Nature}. A monopole component ($n=0$), which can only exist when time reversal (TR) symmetry is broken, yields a linear response~\cite{BerryRev,AnomalRev2010,bohm2003geometric}. A dipole moment ($n=1$) gives rise to a non-linear second-order response~\cite{PRL2015Fu}, which has been measured in ${\rm WTe}_2$~\cite{Ma2019,Kang2019,Min2023} and other materials~\cite{Lee2017,Son2019,Kumar2021,Tiwari2021}. 
Recent studies have probed the quadrupole moment ($n=2$) and its third-order response~\cite{Moore2019,Wang2023,Ye2022,NAg2023,Zhu2023,LawPRB2023,Cano2023}. This has been measured in FeSn, as a transverse voltage generated at three times the frequency of the applied field~\cite{Sankar2023}. Here, we show that the same physics can yield a $\rm dc$ response when a $\rm dc$ applied field is combined with light. 

The Berry curvature and its moments are strongly constrained by the symmetries of the material. We focus on materials that break time reversal (TR) symmetry $K$ and a fourfold rotational symmetry $C_{4}$, while preserving their combination $C_{4}K$. In this case, the Berry curvature monopole and dipole moments must vanish, so that the leading contribution comes from a quadrupole moment. $C_{4}K$ symmetry is realized in altermagnets~\cite{SinovaPRX2022,Sinova2PRX2022,Fedchenko_undated-qm} as well as in magnetically ordered materials belonging to certain magnetic point groups~\cite{LawPRB2023}.
Electronic bands in such materials may be spin-polarized, a feature allowed by broken time-reversal symmetry. We demonstrate that the Hall response will also be generically spin-polarized. 
Intriguingly, the very same symmetry requirements are also invoked in chiral higher order topological crystalline insulators~\cite{Schindler2018}. Here, we restrict our attention to metallic systems where the Hall response appears as a Fermi-surface property~\cite{Haldane2004,Wang2007}. 

Our results provide a counterpoint to the photovoltaic Hall effect. In semiconductor geometries, absorption of circularly polarized light on a biased sample was seen to give rise to spin currents~\cite{bakun1984observation,Ando2010,Sinova2015}. More recently, studies on semi-metallic graphene used a combination of a $\rm dc$ electric field and circularly polarized light to induce a transverse $\rm dc$ current~\cite{Oka2009,Cavalleri2019,Durnev2021}. Circular polarization can be viewed as a TR symmetry-breaking field, akin to the magnetic field in the usual Hall effect. 
Here, we present a logical progression of these ideas to metals with inherent time-reversal breaking such as altermagnets. A Hall current is produced by linearly or elliptically polarized light, but \textit{not} when the polarization is circular. Moreover, this Hall current carries nonzero spin polarization.

\textbf{Results}

\textbf{Framework}

We consider a metal with a single band crossing the Fermi level. We use semiclassical wavepacket dynamics~\cite{BerryRev} to describe transport, denoting wavepacket position as $\bm r$ and momentum as $\bm k$. Following the well-known semiclassical prescription, its velocity is given by $\dot{\bm{r}} = \bm{v} -  \dot{\bm{k}} \times \bm{\Omega}$, where $\bm{v}$ is the group velocity (we use units in which $\hbar = 1$). An anomalous velocity contribution arises from the Berry curvature ${\bm \Omega}$~\cite{bohm2003geometric,BerryRev,AnomalRev2010}. Neglecting external magnetic fields, momentum evolves according to $\dot{\bm{k}} = -e[\bm{E}+\bm{E}_{\rm ac}(t)]$, where the total applied electric field is a combination of a static field $\bm{E}$ and a light-induced field $\bm{E}_{\rm ac}$, as shown in Fig.~\ref{fig:Sample}. The latter is given by $\bm{E}_{\rm ac}(t) = {\rm Re}(\bm{\mathcal E}e^{-i\omega t})$, where $\bm{\mathcal E}$ is a complex vector amplitude. 

Following the Boltzmann transport paradigm, we define a distribution function $f(\bm k,t)$ (see Methods). Treating the applied electric fields as perturbations, we expand this function in powers of the fields up to third order: $f = f_0 + f_1 + f_2 + f_3$, where $f_0$ is the equilibrium Fermi-Dirac distribution function. 
We further divide the distribution into static and time-dependent parts. Keeping only the leading harmonic, we have
\begin{align}
    f_a = f^{(0)}_a + f_a^{(\omega)} e^{-i\omega t} + f_a^{(\omega),*} e^{i\omega t}\quad (a=1,2,3).
\end{align}
At first order, $f^{(0)}_1 \propto E$ and $f_1^{(\omega)} \propto \mathcal{E}$. Second-order terms can be written as $f^{(0)}_2 \propto \mathcal{E}\mathcal{E}^*$ and $f_2^{(\omega)} \propto E\mathcal{E}$. At third order, we have $f^{(0)}_3 \propto E\mathcal{E}\mathcal{E}^*$ and $f_3^{(\omega)} \propto \mathcal{E}\mathcal{E}\mathcal{E}^*$. Other contributions, such as $f^{(0)}_2 \propto EE$, do not affect the Hall currents discussed below. Details of the calculations are given in the Supplemental Material~\cite{supplementary}. 

The electric charge ($c$) current density is given by ${\bm j}_c = -e \int_{\bm k} {\dot{\bm{r}}}\,f(\bm k)$. Focusing on the dc current, we have
\begin{align}
    j_{c,i} =& -e \int_{\bm k} \sum_{a=1}^3  v_i f_a^{(0)} + e^2 \int_{\bm k} \sum_{a=1}^3 \Omega_{ij} E_j  f_a^{(0)} \nonumber\\
&+e^2 \int_{\bm k} \sum_{a=1}^3 \Omega_{ij} [{\mathcal E}_j f_a^{(\omega),*} + c.c.], 
\label{Eq:currentJ0}
\end{align}
where $i,j=x,y$. 
We consider the setup in Fig.~\ref{fig:Sample}, with a static electric field applied along y and Hall current measured along x. As argued in the supplement~\cite{supplementary}, only one component of the Berry curvature, $\Omega_{xy}(\bm k)$, contributes to this current. Henceforth, we disregard other components and denote $\Omega_{xy}$ as simply $\Omega$.

We consider the possibility that this current may carry spin polarization. In materials that break time-reversal symmetry, Bloch states may have unequal weights in the two spin components~\cite{SinovaPRX2022,Krempasky2024}.
To incorporate this effect, we introduce a spin-polarization function, $s(\bm{k})$, 
that represents the expectation value of the $z$ component of spin in the Bloch state at momentum $\bm{k}$. Here, the $z$ direction represents the $C_4K$ axis.
Due to $C_{4}K$ symmetry, $s(\bm{k})$ must switch sign under a four-fold rotation. The spin-polarized ($s$) component of charge current is given by
$\bm {j}_{s} =-e \int_{\bm k} {\dot{\bm {r}}} s(\bm {k}) f(\bm k)$. Henceforth, we refer to this simply as ``spin current''. It can be expressed in terms of the Fermi-Dirac distribution as in Eq.~(\ref{Eq:currentJ0}), by simply including a factor of $s(\bm{k})$ in the integrands.

\begin{figure}
\centering
\includegraphics[width=0.954\linewidth]{Fig 1.pdf}
\caption{Proposed configuration for the light-induced Hall effect. The sample is assumed to be in the $xy$ plane with a fourfold axis along $z$. A static electric field is imposed along $y$. Light, linear or elliptically polarized, impinges from the $z$ axis. The dc current is measured along $x$. }
\label{fig:Sample}
\end{figure}

\textbf{Symmetry considerations} 

We consider a material that breaks $C_{4}$ and $K$, but preserves $C_{4}K$. Note that we will also have $C_{2}$ symmetry, by applying $C_{4}K$ twice.
We consider the geometry shown in Fig.~\ref{fig:Sample} and calculate the dc Hall current to third order in applied electric fields. Assuming $\omega \tau$ is small, we obtain
\begin{align}
   j_{c,x} =& j^{Q}_{c,x} + j^{D}_{c,x},
   \label{Eq:dc-Hall-current}
\end{align} 
where
\begin{align}
   j^{Q}_{c,x} = e^4\tau^2 \Big(\int_{\bm k} f_0~\partial_i\partial_j \Omega \Big) E_y \mathcal{E}_i \mathcal{E}^*_j
   \label{Eq:currentQ}
\end{align} 
represents the contribution from the Berry curvature quadrupole, defined as $\mathcal{Q}^{c}_{ij} = \int_{\bm k}f_0\partial_i \partial_j \Omega$, a Fermi-surface property~\cite{Haldane2004,Wang2007}. 
The second term in Eq.~(\ref{Eq:dc-Hall-current}),
\begin{align}
   j^{D}_{c,x} =-e^4\tau^3 \Big( \int_{\bm k} v_x  \partial_i \partial_j \partial_y f_0 \Big) E_y \mathcal{E}_i \mathcal{E}^*_j, 
  \label{Eq:currentD}
\end{align} 
is a Drude-like contribution. 

Expressions for the current responses crucially depend on the symmetries of the system. In particular, a Berry monopole contribution is ruled out by the $C_{4}K$ symmetry, whereas the linear Drude and Berry dipole contributions vanish due to the $C_2$ symmetry~\cite{LawPRB2023}. Therefore, the leading contributions are given by Eqs.~(\ref{Eq:currentQ}) and (\ref{Eq:currentD}). Any further contribution is either at higher order in electric fields or is scaled down by factors of $\omega\tau$.  

The light-induced Hall current generically carries spin. As with the charge Hall current, the spin Hall current contains two contributions,
\begin{align}
   j_{s,x} =   j^{Q}_{s,x} + j^{D}_{s,x},
   \label{Eq:CurrentS}
\end{align}
where $j^{Q}_{s,x}$ and $j^{D}_{s,x}$ are the Berry-curvature-driven and Drude-like contributions respectively. These contributions are generically nonzero. 


\textbf{Role of polarization}

We assume normal incidence of light on the sample with a spot size much larger than the electronic mean free path. 
We treat light as a spatially  uniform oscillating electric field. Allowing for an arbitrary polarization, the electric field of light of amplitude ${\mathcal E}_{\rm ac}$ is described by
\begin{align}
   {\mathcal E}_x = {\mathcal E}_{\rm ac} \cos\theta,\quad {\mathcal E}_y ={\mathcal E}_{\rm ac} e^{i\phi}\sin\theta. 
  \label{Eq:currentpol}
\end{align}
Linear polarization corresponds to $\phi=0$ or $\pi$. Linear polarization along $x$ ($y$) corresponds to $\theta = 0$ or $\pi$ ($\pi/2$ or $3\pi/2$) with an arbitrary value of $\phi$. When $|{\mathcal E}_x|=|{\mathcal E}_y|$ and $\phi = \pm \pi/2$, the light is circularly polarized. For generic values of $\theta$ and $\phi$, we have elliptical polarization. 

With the ac electric field given by Eq.~(\ref{Eq:currentpol}), 
we now calculate each contribution to the current using the $C_4K$ symmetry. The Berry quadrupole contribution to the charge Hall current takes the form,
\begin{align}
    j^{Q}_{c,x} = 2 e^4\tau^2 E_y \mathcal{E}_{\rm ac}^2 \left(\mathcal{Q}^{c}_{xx}\cos 2\theta  + \mathcal{Q}^{c}_{xy}\sin 2 \theta \cos \phi \right).
\label{eq.currentlight}
\end{align}
Here, 
$\mathcal{Q}_{ij}^{c}$ is the same as the Berry quadrupole density defined earlier. 
The Berry quadrupole contribution to the spin Hall current is given by 
\begin{align}
    j^{Q}_{s,x} = 2 e^4\tau^2 E_y \mathcal{E}_{\rm ac}^2 \mathcal{Q}^{s}_{xx},
\label{eq.scurrentlight}
\end{align}
where we have defined the spin-polarized Berry quadrupole density as $\mathcal{Q}_{ij}^{s}= \int_{\bm k}\,s(\bm k)\, \Omega \partial_i \partial_jf_0$. 
For the charge current arising from third-order Drude contributions, we obtain
\begin{align}
  j_{c, x}^D = -4e^4\tau^3  E_y  \mathcal{E}_{\rm ac}^2 \left[ \mathcal{M}^c_{xxy} \cos 2 \theta + \mathcal{M}^c_{xyy} \sin 2 \theta \cos \phi  \right], 
  \label{eq.jchargeD}
\end{align}
where $\mathcal{M}^c_{ijk} = \int_{\bm k} v_x  \partial_{i} \partial_{j} \partial_{k} f_0$. 
The Drude spin current is given by
\begin{align}
  j_{s, x}^D = -4e^4\tau^3  E_y  \mathcal{E}_{\rm ac}^2 \mathcal{M}^s_{xxy}, 
  \label{eq.scurrentD}
\end{align}
where $\mathcal{M}^s_{ijk} = \int_{\bm k} v_x s(\bm k) \partial_i \partial_j \partial_k f_0 $. 
Expressions (\ref{eq.currentlight}), (\ref{eq.scurrentlight}), (\ref{eq.jchargeD}) and (\ref{eq.scurrentD}) have been derived using the constraints placed by the $C_4K$ symmetry on the response integrals. This symmetry ensures that $\mathcal{Q}_{xx}^{c} = - \mathcal{Q}_{yy}^{c}$, $\mathcal{Q}_{xx}^{s} = \mathcal{Q}_{yy}^{s}$, $Q_{xy}^s = 0$, $\mathcal{M}^c_{xxy} = -\mathcal{M}^c_{yyy}$, $\mathcal{M}^s_{xxy} = \mathcal{M}^s_{yyy}$, and finally $\mathcal{M}^s_{xyy}=0$ (Sec. S4 B in Supplemental Material~\cite{supplementary}). 

Remarkably, as seen from Eqs.~(\ref{eq.currentlight}) and (\ref{eq.jchargeD}), both contributions to charge current vanish for circular polarization (e.g., when $\theta = \pi/4$ and $\phi=\pi/2$). 
The spin current contributions in Eqs.~(\ref{eq.scurrentlight}) and (\ref{eq.scurrentD}) are independent of $\theta$ and $\phi$. They do not change as the polarization of light is varied.

\begin{figure}
\centering
\includegraphics[width=0.58\linewidth]{Fig 2.pdf}
\caption{Proposed tight-binding model for a generic $C_4 K$ material. We have standard hopping processes between nearest neighbours. Along diagonals 
we have Rashba-like hopping shown as blue dotted lines. Altermagnetic order is captured in two spin-dependent hopping processes: $J_1$ between nearest neighbours (red dashed lines) and $J_2$ between next nearest neighbours (black dashed lines). } 
\label{fig:Tight-sample}
\end{figure}

\textbf{Model and Hamiltonian; $C_{4}K$ symmetry}

As a concrete demonstration, we build a minimal model with the required symmetries. We consider a 2D system with two bands arising from electron spin. We begin with a long-wavelength description, followed by a tight-binding model. 

At momenta that are invariant under $C_4K$, the two bands must necessarily be degenerate, giving rise to Dirac points. In the vicinity of each Dirac point, we may write the two-level Hamiltonian as  
$\hat H(\bm k)=d_0({\bm k})\hat\sigma_0+ {\bm d}(\bm k)\cdot\hat{\bm\sigma}$, where $\hat{\bm\sigma}$ is a vector of the Pauli matrices that act on the spin degree of freedom, $\bm k$ is measured from the Dirac point, and the coefficients $d_i(\bm k)$ are real functions of $\bm k$. 
With the symmetry group generated by the antiunitary operation $C_{4}K$, at small $\bm k$ we have~\cite{supplementary}
\begin{eqnarray}
   d_0(\bm k) &=& a_0 + a_1( k^2_x + k^2_y ),\nonumber\\
     d_1(\bm k) &=& b_1 k_x+b_2 k_y, \nonumber\\
     d_2(\bm k) &=&-b_2 k_x + b_1 k_y , \nonumber\\ 
     d_3(\bm k) &=& m_1 (k_x^2-k_y^2) + 2m_2 k_xk_y.
    \label{Eq:ConstraintsSpin}
\end{eqnarray}
We have used the lowest-order polynomial expressions for the components of $\bm d$. 
Similar expressions have previously been used as toy models for altermagnets~\cite{MacDonald2022}.
The Berry curvature can be immediately computed using 
$\Omega = \epsilon^{ijk} d_i \left(\partial_{k_x} d_j \partial_{k_y} d_k\right)/{2|\bm d|^3}$, where $|\bm d| = \sqrt{d_1^2+d_2^2+d_3^2}$. 
This results in a quadrupole-like distribution of $\Omega$ around $\bm k=0$~\cite{supplementary}.
In the neighbourhood of each Dirac point, we will have a long-wavelength Hamiltonian of the form given in Eq.~(\ref{Eq:ConstraintsSpin}).  

To better understand the origin of various terms in the Hamiltonian, we construct a tight-binding model that reproduces Eq.~(\ref{Eq:ConstraintsSpin}). 
We start with a 2D square lattice as shown in Fig.~\ref{fig:Tight-sample}, with the lattice constant set to unity.  We expect to find Dirac points at two $C_4K$-invariant momenta: $\bm k=(0,0)$ and $(\pi,\pi)$, corresponding to $\Gamma$ and $M$ points in the Brillouin zone respectively.   
Apart from the usual hopping processes, we have spin-dependent hopping between nearest neighbours, which could arise from the Rashba spin-orbit coupling. We introduce ``altermagnetic order parameters'' $J_1$ and $J_2$, which encode preferential hopping of each spin along nearest and next-nearest neighbour bonds. Crucially, the $J_1$ and $J_2$ processes break $C_4$ and $K$, but preserve $C_4 K$. We arrive at the following Hamiltonian in momentum space:
\begin{align}
    \hat H(\bm k) =& -t \big(\cos k_x   + \cos  k_y  \big) \hat\sigma_0 \nonumber\\ 
    & +\frac{\lambda}{2} \left[\sin  \left(k_x +k_y \right)\hat\sigma_x + \sin \left(k_y-k_x \right) \hat\sigma_y \right] \nonumber\\ 
  &+ \left[ J_1(\cos  k_x  - \cos  k_y ) + J_2 \sin  k_x  \sin k_y \right]\hat\sigma_z,
\label{Eq:HamiltonianMomentum}
\end{align}
where $t$ is the usual hopping parameter and $\lambda$ corresponds to the Rashba spin-orbit coupling.

\begin{figure}
\centering
\includegraphics[width=0.995\linewidth]{Fig 3}
\caption{An example of band structure and Fermi surface pockets. (a) Band structure and (b) Fermi pockets for $\lambda=0.4$,  $\mu=-0.15$, $J_1=J_2=1$, and $t=0.02$ in arbitrary units of energy.  }
\label{Fig:bands}
\end{figure}

The resulting band structure is plotted in Fig.~\ref{Fig:bands}, for a certain choice of model parameters. 
Changing the chemical potential affects the shape of the Fermi pockets (Supplemental Material, Figs. S1 and S2~\cite{supplementary}). To proceed analytically, we examine the behavior of the system near the Dirac nodes. Near the $\Gamma$ point, we recover the long-wavelength form of Eq.~(\ref{Eq:ConstraintsSpin}) with 
$a_0 =-2t $, $a_1=t/2$, $b_1 = b_2 =\lambda /2$, $m_1 =-J_1/2$, $m_2 =J_2/2$. Near the $M$ point, we recover the same form, but with  $a_0 =2t$, $a_1=t/2$, $b_1 = b_2 = \lambda/2$, $m_1 =J_1/2$, and $m_2=J_2/2$. 

At each Dirac node, we have an upper band and a lower band. Their Berry curvatures are given by 
\begin{align}
   \Omega^{\eta}_{\pm} =  \pm \eta \frac{ [J_1 (k_x^2 - k_y^2) -2 \eta J_2  k_x k_y] \lambda^2  }{\left( [J_1  (k_x^2 - k_y^2)-2\eta J_2  k_xk_y]^2 +2 \lambda^2  k^2 \right)^{3/2}} ,
\end{align}
where $\eta =+1$ around $\Gamma$ and $-1$ around $M$. The upper (lower) sign applies for the upper (lower) band. 
Using these expressions, we may evaluate the quadrupole densities, $\mathcal{Q}_{ij}^{c,s}$ , which appear in the charge and spin Hall currents, see Eqs.~(\ref{eq.currentlight}) and (\ref{eq.scurrentlight}). Notably, the $J_1$ and $J_2$ order parameters produce distinct quadrupole patterns. The former can be viewed as creating a $d_{x^2-y^2}$-wave altermagnet, whereas the latter yields a $d_{xy}$-wave altermagnet (Fig. S3 in the Supplemental Material~\cite{supplementary}).

On examining the Bloch states of each band, we find them to be generically spin-polarized. The entries in each eigenspinor have unequal amplitudes. This allows us to define the spin-polarization function 
\begin{align}
   s^{\eta}_{\pm}(\bm k) = \mp\eta \frac{J_1   (k^2_x-k^2_y) - 2\eta J_2   k_x k_y }{\sqrt{ [J_1   ( k_x^2 - k_y^2 ) - 2\eta J_2   k_x k_y]^2 + 2 \lambda^2   k^2}},
\end{align}
which enters the spin Hall current in Eqs.~(\ref{Eq:CurrentS}), (\ref{eq.scurrentlight}) and (\ref{eq.scurrentD}).

\textbf{Light-induced Hall currents}

As an explicit demonstration, we calculate light-induced Hall current with the following simplifying assumptions. The Fermi energy is taken to be close to both Dirac points, resulting in two small Fermi pockets.
We further assume weak altermagnetic order with $J_1,J_2 \ll \lambda$. This results in nearly circular Fermi surfaces (Supplemental Material~\cite{supplementary}). Below, we calculate current contributions to leading order in $J_{1,2}  /\lambda$. For concreteness, we suppose that the Fermi energy crosses the lower band at each pocket.  Neglecting inter-pocket scattering, the charge and spin currents acquire contributions from each pocket separately.  

 The charge current induced by the Berry curvature quadrupole is obtained from Eq.~(\ref{eq.currentlight}), including contributions from both Fermi pockets:
\begin{align}
    j^{Q}_{c,x} =&\frac{ e^4\tau^2}{16 \pi }  E_y {\mathcal E}_{\rm ac}^2 \bigg[J_1   \left( \frac{1}{|\epsilon_{\Gamma} - \mu|} - \frac{1}{|\epsilon_{M} - \mu|} \right) \cos 2\theta \nonumber\\ 
    &+ J_2  \left( \frac{1}{|\epsilon_{\Gamma} - \mu|} - \frac{1}{|\epsilon_{M} - \mu|} \right) \sin 2\theta \cos \phi \bigg],
\label{eq.jcQmodel}
\end{align}
where $\epsilon_{\Gamma}$ and $\epsilon_{M}$ are the energies of the degenerate bands at the $\Gamma$ and $M$ points, respectively. $\mu$ is the chemical potential. 
The quadrupole-induced spin current of Eq.~(\ref{eq.scurrentlight}) takes the form  
\begin{align}
    j^{Q}_{s,x} = \frac{ e^4\tau^2}{64 \pi}  \frac{(J_1^2+J_2^2)^2  }{\lambda^4}E_y {\mathcal E}_{\rm ac}^2 \bigg[\frac{|\epsilon_{\Gamma} - \mu|^2}{\lambda^2} + \frac{|\epsilon_{M} - \mu|^2}{\lambda^2} \bigg].
\end{align}
These expressions show that the quadrupole-induced Hall currents arise from the altermagnetic order parameters, $J_1$ and $J_2$. If these order parameters were to vanish, so would the Berry-curvature quadrupole and its contribution to the dc Hall current. 

We next calculate the Drude current by including contributions from the two Fermi pockets.  The Drude charge current of Eq.~(\ref{eq.jchargeD}) comes out to be
\begin{align}
  j^{D}_{c,x} =&\frac{ e^4\tau^3 \lambda^2  }{8 \pi} E_y {\mathcal E}_{\rm ac}^2\bigg[ 
  \left( \frac{1}{|\epsilon_\Gamma - \mu|} + \frac{1}{|\epsilon_{M} - \mu|} \right) \sin 2 \theta \cos \phi
  \nonumber\\
  &+\frac{3}{2} \frac{J_1J_2 }{\lambda^4}\big(|\epsilon_\Gamma - \mu| + |\epsilon_{M} - \mu|  \big) \cos 2\theta 
  \bigg], 
\end{align} 
whereas the Drude spin current of Eq.~(\ref{eq.scurrentD}) yields 
\begin{align}
  j^{D}_{s,x}  =\frac{ e^4\tau^3 J_2  }{2 \pi} E_y {\mathcal E}_{\rm ac}^2 .
  \label{eq.jsDmodel}
\end{align}
The Supplemental Material provides estimates of these current magnitudes for realistic parameters~\cite{supplementary}.

\textbf{Discussion}

We have demonstrated a light-induced Hall current in materials with $C_4K$ symmetry. This current carries a spin polarization that can be tuned by varying the polarization of light. This result, expressed in Eqs.~(\ref{Eq:dc-Hall-current}-\ref{eq.scurrentD}), arises purely from $C_4K$ symmetry and is applicable to both 2D as well as 3D materials, e.g., to thick films where light can penetrate uniformly.
We emphasize two features: (i) the charge Hall current vanishes when light is circularly polarized, and (ii) the spin current does not vary with polarization. As an explicit demonstration, we have discussed a 2D toy model where  analytic forms are derived. Our analysis is based on the assumption that frequency is lower than the inverse relaxation time ($\omega\tau \ll 1$). With relaxation times typically ranging from femtoseconds to picoseconds in metals, infrared or microwave radiation may be used to see the predicted phenomena. 

Our results can be readily tested by shining light on altermagnetic materials such as $\rm{RuO_2}$, $\rm{MnO_2}$ and $\rm{MnF_2}$~\cite{Sinova2PRX2022,SinovaPRX2022}. Light-induced currents complement various other transport properties known in altermagnets~\cite{Bose2022,Sun2023,Das2023,Bai2023,Zhou2024}. They  may provide a simpler signature of $C_4K$ symmetry, with a dc charge response that does not require a spin-sensitive apparatus. Our results can also apply to other systems with Berry curvature quadrupoles such as metals with ferro-octupolar order \cite{Sorn2024}. In any such material, light can be used as a switch to generate currents on demand. This approach can complement other architectures for switchable spin currents~\cite{Torres2014,Qiu2018,Zhang2022}.

\textbf{Methods}

The distribution function $f(\bm k,t)$ satisfies the Boltzmann kinetic equation
\begin{align}
  \frac{\partial f}{\partial t}  + \dot{\bm k}\,\cdot \frac{\partial f}{\partial\bm k} = -\frac{f-f_0}{\tau}.
  \label{Eq:Bolzmann-eq}
\end{align}
The relaxation time $\tau$ is assumed to be a constant for simplicity. We use the semi-classical equations for $\dot{\bm k}$ and $\dot{\bm r}$ and solve for $f(\bm k, t)$ as a series expansion in the electric fields to calculate currents.

\textbf{Data availability}

The data supporting the findings of this study are available from the 
authors upon reasonable request.

\textbf{Acknowledgments}

We thank R. Shankar, A. Soori, S. A. Jafari, R. Ghadimi, and K. Farain for helpful discussions. This work was supported by the Natural Sciences and Engineering Research Council of Canada through Discovery Grants 2022-05240 (RG) and 2021-03705 (KS).

\textbf{Author contributions}

TF, RG, and KS conceived and designed the research project. TF carried out the calculations. All authors contributed equally to the writing and editing of the manuscript.

\textbf{Competing interests}

The authors declare no competing financial interests.

\bibliography{Ref2D}

\newpage \clearpage
 
\onecolumngrid 
\setcounter{secnumdepth}{3}
\renewcommand{\theequation}{S\arabic{equation}}
\renewcommand{\thefigure}{S\arabic{figure}}
\begin{center}
 \textbf{\large Supplemental Material for ``Light-induced charge and spin Hall currents in materials with $C_4K$ symmetry''}\\[.5cm]
T. Farajollahpour,$^1$ R. Ganesh,$^1$ and K. V. Samokhin$^1$\\[.4cm]
{\itshape ${}^1$Department of Physics, Brock University, St. Catharines, Ontario L2S 3A1, Canada\\}
(Dated: \today)\\[2cm]
\end{center}
\section{Semiclassical equations of motion}

In the semiclassical description of transport, the electric current density can be expressed in terms of the distribution function $f({\bm k})$ as 
\begin{align}
     j_i =-e \int_{\bm k} {\dot{r}}_i f(\bm k), 
     \label{Eq:current}
\end{align}
where $\int_{\bm k} \equiv \int d^d{\bm k}/(2\pi)^d$ in $d$ dimensions and ${\dot{\bm r}}$ is the velocity of a wavepacket. The velocity obeys the semiclassical equations of motion, modified by the Berry curvature~\cite{BerryReview,AnomalRev2010,bohm2003geometric,shun2018topological}, 
\begin{align}
\dot{r}_i = v_i + \Omega_{ij} \dot{k}_j,
\end{align}
and
\begin{align}
 \dot{k}_i = -e [E_i+E_{{\rm ac},i}(t)].
   \label{Eq:EQMsup}
\end{align}
Here, $\Omega_{ij}$ is the Berry curvature and $\bm v$ is the group velocity. The total applied electric field is a combination of a static field $\bm{E}$ and a light-induced field $\bm{E}_{\rm ac}$. The latter is given by $\bm{E}_{\rm ac}(t) = {\rm Re}(\bm{\mathcal E}e^{-i\omega t})$, where $\bm{\mathcal E}$ is a complex vector. In the relaxation time approximation, the electron distribution function follows the Boltzmann equation,
\begin{align}
   \frac{\partial f}{ \partial t}  + \dot{k}_i \frac{\partial f}{\partial k_i} = - \frac{f-f_0}{\tau},
   \label{Eq:Boltzman}
\end{align}
where $f_0$ represents the equilibrium Fermi distribution function and $\tau$ denotes the relaxation time, assumed to be a constant for simplicity. By inserting Eq.~(\ref{Eq:EQMsup}) we have
\begin{align}
   \frac{\partial f}{ \partial t}  -e  [E_i+E_{{\rm ac},i}(t)] \frac{\partial f}{\partial k_i} = -\frac{f-f_0}{\tau}.
   \label{Eq:Boltz}
\end{align}

Treating the applied electric fields as perturbations, we expand the solution of Eq. (\ref{Eq:Boltz}) in powers of the fields up to third order:
\begin{align}
    f = f_0 + f_1 + f_2 + f_3. 
\end{align} 
We further divide the distribution into static and time-dependent parts. Keeping only the leading harmonic, we have
\begin{align}
    f_a = f^{(0)}_a + f_a^{(\omega)} e^{-i\omega t} + f_a^{(\omega),*} e^{i\omega t}\quad (a=1,2,3),
    \label{eq.fs}
\end{align}
where $a$ represents the order of the expansion in electric fields. 
We substitute the expansion of Eq.~(\ref{eq.fs}) into Eq.~(\ref{Eq:Boltz}) and solve for each component of $f$ using an order-by-order approach, proceeding sequentially from lower to higher orders. At first order, $f^{(0)}_1 \propto E$ and $f_1^{(\omega)} \propto \mathcal{E}$. Second-order terms can be written as $f^{(0)}_2 \propto \mathcal{E}\mathcal{E}^*$ and $f_2^{(\omega)} \propto E\mathcal{E}$. At third order, we have $f^{(0)}_3 \propto E\mathcal{E}\mathcal{E}^*$ and $f_3^{(\omega)} \propto \mathcal{E}\mathcal{E}\mathcal{E}^*$. We neglect certain contributions, such as $f^{(0)}_2 \propto EE$, that do not impact the Hall current. We find
\begin{align}
  &  f_1^{(0)} = e \tau E_i \partial_i f_0, \nonumber\\ 
   & f_1^{(\omega)} = \frac{e \tau \mathcal{E}_i}{1-i\omega\tau} \partial_i f_0, \nonumber\\ 
   & f^{(0)}_2 = e^2 \tau^2 \left(\frac{\mathcal{E}^*_i \mathcal{E}_j}{1-i\omega \tau } + \frac{\mathcal{E}_i \mathcal{E}^*_j}{1+i\omega \tau } \right) \partial_i \partial_j f_0, \nonumber\\
   & f_2^{(\omega)} = e^2 \tau^2  \left(\frac{E_i \mathcal{E}_j}{(1-i\omega \tau)^2} + \frac{E_j \mathcal{E}_i}{1-i\omega \tau} \right) \partial_i \partial_j f_0,\nonumber\\
      & f_3^{(0)} = e^3 \tau^3 \left( \frac{\mathcal{E}_i^* \mathcal{E}_j E_k}{(1-i \omega \tau)^2}+ \frac{\mathcal{E}_i^* \mathcal{E}_k E_j}{1-i \omega \tau} + \frac{\mathcal{E}_i \mathcal{E}_j^* E_k}{(1+i \omega \tau)^2}+ \frac{\mathcal{E}_i \mathcal{E}_k^* E_j}{1+i \omega \tau} \right) \partial_i \partial_j \partial_k f_0, \nonumber\\
    &  f_3^{(\omega)} = e^3\tau^3 \left(\frac{\mathcal{E}_i \mathcal{E}^*_j \mathcal{E}_k}{(1-i\omega \tau)(1-i\omega\tau)} + \frac{\mathcal{E}_i \mathcal{E}_j \mathcal{E}^*_k}{(1+i\omega \tau)(1-i\omega\tau)}\right) \partial_i \partial_j \partial_k f_0.
\label{Eq:Finalf}
\end{align}

\section{dc current}
The dc current charge density is given by the expression
\begin{align}
    j^{(0)}_i = -e \int_{\bm k} \sum_{n}  v_i f_n^{(0)} + e^2 E_j \int_{\bm k} \sum_{n} \Omega_{ij}   f_n^{(0)} + e^2 \int_{\bm k} \sum_{n} \Omega_{ij} [\mathcal{E}_j f_n^{(\omega),*} + c. c.]. 
    \label{Eq:J0}
\end{align}
Substituting the explicit solutions (\ref{Eq:Finalf}) into Eq.~(\ref{Eq:J0}), we arrive at 
\begin{align}
   j^{(0)}_i &=-e \int_{\bm k} v_i f_0 +e^2 E_j\int_{\bm k} \Omega_{ij}  f_0-e^2 \tau E_j\int_{\bm k} v_i \partial_j f_0  +e^3 \tau E_j E_k\int_{\bm k} \Omega_{ij}\partial_k f_0 \nonumber\\
   &+\biggl\{\frac{e^3\tau}{1+i\omega \tau} \mathcal{E}_j \mathcal{E}_k^*\left[ \int_{\bm k} \Omega_{ij} \partial_k f_0 - e \tau\int_{\bm k} v_i \partial_j \partial_k f_0  \right] + c.c. \biggr\} \nonumber\\ 
  & +\biggl\{ \frac{e^4\tau^2}{1+i\omega \tau}  E_j \mathcal{E}_k \mathcal{E}^*_m\left[
  \int_{\bm k} \Omega_{ij}  \partial_m\partial_k f_0 ~  -\frac{\tau}{1+i\omega \tau} \int_{\bm k} v_i \partial_k \partial_m \partial_j f_0  -
  \tau \int_{\bm k} v_i \partial_k \partial_j \partial_m f_0  \right] + c.c. \biggr\} .
  \label{eq.currentj}
\end{align}
The term
$D_{k;ij} = \int_{\bm k} f_0~(\partial_k \Omega_{ij}) = -  \int_{\bm k}   \Omega_{ij} \partial_k f_0$
is referred to as the Berry curvature dipole~\cite{PRL2015Fu}, while
\begin{align}
    Q_{k m;ij} = \int_{\bm k} f_0 (\partial_k \partial_m \Omega_{ij}) =   \int_{\bm k}   \Omega_{ij} \partial_k \partial_m f_0
\end{align}
is called the Berry curvature quadrupole~\cite{Moore2019,LawPRB2023}. In 2D materials, only one component of the Berry curvature, $\Omega_{xy}$, is non-zero. In 3D materials, with our setup, with a static electric field along y and Hall current measured along x, only $\Omega_{xy}$ contributes to the current. This can be seen by setting $i=x$ and $j=y$ in Eq.~(\ref{eq.currentj}) above. For brevity, we neglect all other components of Berry curvature in the expressions below. As discussed in the main text, the symmetries in the problem impose some constraints on the integrals in Eq.~(\ref{eq.currentj}). With the $C_4K$ symmetry, the Hall current (perpendicular to the static field) will only receive contributions at third order in the electric field. We focus on these terms in the following sections.
\subsection{Charge current}
We now present explicit expressions for the current contributions that arise from various combinations of electric fields. We assume that the static field is along $y$, i.e. ${\bm E}=E_y\hat{\bm y}$, whereas the ac field can have both $x$ and $y$ components. Below, we list the third-order terms proportional to 
$E_y\mathcal{E}_y\mathcal{E}_y$, $E_y \mathcal{E}_x \mathcal{E}_x$ and $E_y\mathcal{E}_x \mathcal{E}_y$, from which the final expressions in the main text are obtained. 

The charge current along the $x$ direction, $j^{(0)}_x$, contains the following third-order contributions:
\begin{align}
    & \nonumber |\mathcal{E}_y|^2  E_y\bigg\{\frac{e^4 \tau^2}{1+i\omega \tau} \bigg[  \int_{\bm k} \Omega_{xy}  \partial_y\partial_y f_0     -\frac{{2\tau + i\omega\tau^2}}{1+i\omega \tau} \int_{\bm k} v_x \partial_y \partial_y \partial_y f_0   \bigg] + c.c. \bigg\} \\
    & \nonumber  +|\mathcal{E}_x|^2 E_y\bigg\{\frac{e^4\tau^2}{1+i\omega \tau}\bigg[ \int_{\bm k} \Omega_{xy}  \partial_x\partial_x f_0    -\frac{2\tau + i\omega \tau^2}{1+i\omega \tau} \int_{\bm k} v_x \partial_x \partial_x \partial_y f_0   \bigg] + c.c. \bigg\} \\
    & +(\mathcal{E}_x \mathcal{E}^*_y + c.c.)E_y \bigg\{\frac{e^4\tau^2}{1+i\omega \tau}\bigg[ \int_{\bm k} \Omega_{xy}  \partial_x\partial_y f_0   -\frac{2\tau + i\omega \tau^2}{1+i\omega \tau} \int_{\bm k} v_x \partial_x \partial_y \partial_y f_0   \bigg]+ c.c. \bigg\}.
  \label{Eq:currentEs}
\end{align}
The spin current contains the same integrals, but with an additional factor in each integrand to account for spin polarization. Below, we provide explicit expressions for the spin polarization in a class of models. 

\section{$C_{4}K$ symmetry} 
\label{Sec:BCQ}
\subsection{The Berry curvature under $C_4K$} 
We examine the action of $C_4K$ on the Berry curvature. To do this, we examine the semiclassical equations of motion,
\begin{align}
\dot{x}= v_x + \Omega_{xy} \dot{k}_y,\qquad \dot{y}= v_y - \Omega_{xy} \dot{k}_x. 
\end{align}
Under $C_4K$, we have 
$\dot{x} \rightarrow -\dot{y}, ~v_x \rightarrow -v_y,~ \dot{k}_y \rightarrow -\dot{k}_x$.
In order to preserve the equations of motion, we must have $\Omega_{xy} \rightarrow -\Omega_{xy}$.

\subsection{Symmetry constraints on response functions} 
\label{sec.symmconstraints}

The expressions in Eq.~(\ref{Eq:currentEs}) involve two types of integrals, one that involves the Berry curvature $\Omega\equiv\Omega_{xy}$ and one that does not. They can be written as  
\begin{align}
     Q_{ij}^c =\int_{\bm k} \Omega\, \partial_i \partial_j f_0
\end{align}
and
\begin{align}
   \mathcal{M}^c_{ijk} =  \int_{\bm k} v_x\, \partial_i \partial_j \partial_k f_0.
\end{align}
The superscript ``$c$'' here indicates that these integrals appear in the charge current. We now show that they are strongly constrained by the $C_4K$ symmetry. As an example, let us consider $Q_{xx}^c $,%
\begin{align}
  Q_{xx}^c &=
  \frac{1}{2} \int_{\bm k}  \overbrace{\Omega(\partial_x \partial_x +\partial_y \partial_y)f_0}^{{\text A}}+ \frac{1}{2} \int_{\bm k} \overbrace{\Omega (\partial_x \partial_x -\partial_y \partial_y)  f_0}^{{\text B}}  \nonumber\\ 
  &=\frac{1}{2} \int_{\bm k} \Omega  (\partial_x \partial_x -\partial_y \partial_y)f_0.
  \label{Eq:CQxx}
\end{align}
In the first line, we expressed $Q_{xx}^c$ as the sum of two integrals over the Brillouin zone. In the first term, the integrand (denoted as $A$) is odd under $C_4K$, because $\Omega$ is odd under while $f_0$ is even. As the integrand $A$ is odd, it integrates to zero. In the second term, the integrand (denoted as $B$), is even under $C_4K$ and may integrate to a non-zero value. Along the same lines, we have 
\begin{align}
   Q_{yy}^c   =  \frac{1}{2} \int_{\bm k} \Omega(\partial_x \partial_x +\partial_y \partial_y)f_0 - \frac{1}{2} \int_{\bm k} \Omega(\partial_x \partial_x -\partial_y \partial_y)  f_0 
   =-\frac{1}{2} \int_{\bm k} \Omega (\partial_x \partial_x -\partial_y \partial_y)f_0.
  \label{Eq:CQyy}
\end{align}
From Eqs.~(\ref{Eq:CQxx}) and (\ref{Eq:CQyy}), we conclude that $
Q_{xx}^c = -Q^c_{yy}$. In the same manner, it is easy to see that $\mathcal{M}_{xxy}^c =-\mathcal{M}_{yyy}^c$.

We next consider the spin current, which involves the spin polarization function $s(\bm{k})$ introduced in the main text. Below, in Sec.~\ref{Sec:SpinPol}, we provide expressions for $s(\bm{k})$ that are valid in any two-band model. For now, we note that $s(\bm{k})$ must be odd under $C_4K$. The expressions for the spin current are obtained from Eq.~(\ref{Eq:currentEs}) upon including a factor of $s(\bm{k})$ in the integrals. They involve two quantities, denoted as
\begin{align}
Q_{ij}^s = \int_{\bm{k}} s(\bm{k}) \Omega\, \partial_i \partial_j f_0
\end{align}
and 
\begin{align}
\mathcal{M}^s_{ijk} = \int_{\bm{k}} s(\bm{k}) v_x\, \partial_i \partial_j \partial_k f_0.
\end{align}
We express $Q_{xx}^s $ as the sum of two terms:
\begin{align}
  Q_{xx}^s  &=   \frac{1}{2} \int_{\bm k} s(\bm k) \Omega(\partial_x \partial_x +\partial_y \partial_y)f_0+  \frac{1}{2} \int_{\bm k} s(\bm k) \Omega (\partial_x \partial_x -\partial_y \partial_y)  f_0    \nonumber\\ 
  &= \frac{1}{2} \int_{\bm k} s(\bm k) \Omega  (\partial_x \partial_x +\partial_y \partial_y)f_0.
  \label{Eq:SQxx}
\end{align}
We have dropped the second term which is odd under $C_4K$ and must therefore integrate to zero. Similarly, we have
\begin{align}
 Q_{yy}^s  &=  \frac{1}{2} \int_{\bm k} s(\bm k) \Omega(\partial_x \partial_x +\partial_y \partial_y)f_0-  \frac{1}{2} \int_{\bm k} s(\bm k) \Omega (\partial_x \partial_x -\partial_y \partial_y)  f_0   \nonumber\\ 
  &=  \frac{1}{2} \int_{\bm k} s(\bm k) \Omega  (\partial_x \partial_x +\partial_y \partial_y)f_0.
  \label{Eq:SQyy}
\end{align}
From Eqs.~(\ref{Eq:SQxx}) and (\ref{Eq:SQyy}), we conclude that $Q_{xx}^s  = Q_{yy}^s$. With similar arguments, we find $Q_{xy}^s  =0$, $\mathcal{M}_{xyy}^s=0$, and $\mathcal{M}_{xxy}^s = \mathcal{M}_{yyy}^s$. 

 
\subsection{Polarization dependence of the charge and spin currents} 

Having discussed the constraints imposed by the $C_4K$ symmetry on the response integrals, we put together expressions for the induced Hall current. Assuming $\omega\tau \ll 1$, Eq.~(\ref{Eq:currentEs}) can be written as
\begin{align}
j_{x, \alpha}^Q = 2{e^4\tau^2} \left[Q^{\alpha}_{yy} E_y \mathcal{E}_y \mathcal{E}_y^* + Q^{\alpha}_{xx} E_y \mathcal{E}_x \mathcal{E}_x^* + Q^\alpha_{xy} E_y(\mathcal{E}_y \mathcal{E}^*_x + \mathcal{E}_x \mathcal{E}^*_y)\right].
\end{align}
Here, $\alpha=c,s$ denotes charge ($c$) or spin ($s$) currents. The transverse components of the Drude current are given by
\begin{align}
j_{x, \alpha}^D = -4e^4\tau^3 E_y \left[ \mathcal{M}^\alpha_{yyy} |\mathcal{E}_y|^2 + \mathcal{M}^\alpha_{xxy} |\mathcal{E}_x|^2 + \mathcal{M}^\alpha_{xyy} (\mathcal{E}_y\mathcal{E}^*_x + \mathcal{E}_x\mathcal{E}^*_y)\right].
\end{align}
Within the paraxial approximation~\cite{andrews2011}, the electric fields associated with light are spatially uniform. As described in the main text, for  arbitrary polarization the complex amplitude of the light field can be written as 
\begin{align}
 & \mathcal{E}_x = \mathcal{E}_{\rm ac} \cos \theta,\quad \mathcal{E}_y = \mathcal{E}_{\rm ac} \sin \theta e^{i \phi}. 
 \label{Eq:electricprofile}
\end{align}
Using the $C_4 K$ symmetry ($Q^c_{yy} = - Q^c_{xx}$), we have
\begin{align}
    j^{Q}_{c,x} = 2 e^4\tau^2 E_y \mathcal{E}_{\rm ac}^2 \left[Q^{c}_{xx}\cos 2\theta  + Q^{c}_{xy}\sin 2 \theta \cos \phi \right].
\end{align}
The Drude contribution, with $\mathcal{M}_{xxy}^c =-\mathcal{M}_{yyy}^c$ due to $C_4K$ symmetry, is given by 
\begin{align}
  j_{x, c}^D = -4e^4\tau^3  E_y  \mathcal{E}_{\rm ac}^2 \left[ \mathcal{M}^c_{xxy} \cos 2 \theta+  \mathcal{M}^c_{xyy} \sin 2 \theta \cos \phi  \right]. 
\end{align}
For the spin currents, the $C_4K$ symmetry yields $Q^s_{xx} = Q^s_{yy} $ and $Q_{xy}^s =0$, and we obtain
\begin{align}
    j^{Q}_{s,x} =  2e^4\tau^2 E_y \mathcal{E}_{\rm ac}^2 Q^{s}_{xx}.
\end{align}
For the Drude spin contribution, with $\mathcal{M}^s_{xxy} = \mathcal{M}^s_{yyy}$ and $\mathcal{M}^s_{xyy} = 0$, we obtain 
\begin{align}
  j_{x, s}^D = -4e^4\tau^3  E_y  \mathcal{E}_{\rm ac}^2 \mathcal{M}^s_{xxy}. 
\end{align}

\section{Two-band systems with $C_4 K$ symmetry} 

\subsection{Continuum model}

We derive 
the general ``two-level'' Hamiltonian at long wavelengths in a 2D system with $C_4K$ symmetry.
We take $C_4$ to represent fourfold rotations about the $z$ axis with $K = -i\hat\sigma_y K_0$ being the time reversal operation~\cite{Laxbook}. The generic long-wavelength two-level Hamiltonian can be written as:
\begin{align}
  H = \sum_{\bm k, \alpha} d_0(\bm k)  a^\dagger_{\bm k, \alpha} a_{\bm k, \alpha} + \sum_{\bm k, \alpha \beta}  \bm d (\bm k)\cdot {{\bm \sigma}}_{\alpha \beta}~ a^\dagger_{\bm k, \alpha} a_{\bm k, \beta},  
  \label{suppl-eq:Hamiltonian}
\end{align}
where $\alpha,\beta = \uparrow,\downarrow$ is the spin projection, $d_0$ and $\bm d$ are real functions of $\bm k$, and ${\hat {\bm \sigma}}$ is the vector of the Pauli matrices. The operators $a^\dagger_{\bm k, \alpha}$ ($a_{\bm k, \alpha}$) create (annihilate) electrons in the pure spin states
\begin{align}
    \langle \bm r | \bm k, \uparrow \rangle = \frac{1}{\sqrt{\mathcal{V}}} e^{i \bm k \cdot \bm r}\left(\begin{matrix} 1 \\ 0 \end{matrix}\right),\qquad
    \langle \bm r | \bm k, \downarrow \rangle = \frac{1}{\sqrt{\mathcal{V}}} e^{i \bm k \cdot \bm r}\left(\begin{matrix} 0 \\ 1 \end{matrix}\right),
\label{suppl-eq:pure-spinors}
\end{align}
where ${\cal V}$ is the system volume. 

Under the operation $g=C_4 K$, the wave functions (\ref{suppl-eq:pure-spinors}) transform as $g|\bm k,\alpha\rangle= \hat{D}^{(1/2)}(C_4)K |\bm k,\alpha\rangle$, where $\hat D^{(1/2)}(R)$ is the spinor representation of a rotation $R$ \cite{Laxbook}. Therefore, $g|\bm k, \uparrow \rangle = e^{i\pi/4} |-C_4 \bm k, \downarrow \rangle$ and $g| \bm k, \downarrow \rangle = -e^{-i\pi/4} |-C_4 \bm k, \uparrow \rangle$, and we obtain the following transformation rules for the electron creation and annihilation operators:
\begin{align}
    g (f a^\dagger_{\bm k,\uparrow} ) g^{-1} = f^*e^{i\pi/4} a^\dagger_{-C_4\bm k,\downarrow},\qquad 
    g (f a^\dagger_{\bm k,\downarrow} ) g^{-1} = -f^*e^{-i\pi/4} a^\dagger_{-C_4\bm k,\uparrow},
\label{suppl-eq:g-transform}
\end{align}
where we introduced a $c$-number constant $f$ to emphasize the antilinearity of $g$. 
Using Eq. (\ref{suppl-eq:g-transform}) in the invariance condition $gHg^{-1}=H$ for the Hamiltonian (\ref{suppl-eq:Hamiltonian}), we obtain the following constraints on the coefficients:
\begin{equation}
    d_0(\bm k)=d_0(-C_4^{-1}\bm k),\quad
    d_x(\bm k)=d_y(-C_4^{-1}\bm k),\quad
    d_y(\bm k)=-d_x(-C_4^{-1}\bm k),\quad
    d_z(\bm k)=-d_z(-C_4^{-1}\bm k).
\label{suppl-eq:C4-constraints}
\end{equation}
In a similar fashion, using the invariance of the Hamiltonian under $g^2=(C_4 K)^2$ and the fact that 
$C_{2z}{\bm k}=-{\bm k}$ in 2D, we obtain that $d_0$ and $d_z$ are even in ${\bm k}$, while $d_x$ and $d_y$ are odd. Introducing $d_{\pm}=d_x\pm id_y$ and $k_\pm=k_x\pm ik_y$, Eq. (\ref{suppl-eq:C4-constraints}) takes the form
\begin{equation}
    d_0(k_+,k_-) =  d_0 (ik_+,-ik_-), \quad
    d_\pm(k_+,k_-) = \mp i d_\pm (ik_+,-ik_-), \quad 
    d_z(k_+,k_-) = - d_z (ik_+,-ik_-).
\end{equation}
The lowest-order polynomial in ${\bm k}$ solutions of these constraints can be written as
\begin{align}
    &  d_0 (\bm k) = a_0 + a_1 (k_x^2+k_y^2), \nonumber\\ 
    &  d_x (\bm k) = b_1 k_x + b_2 k_y, \nonumber\\ 
    &  d_y (\bm k) =- b_2 k_x + b_1 k_y, \nonumber\\ 
    &  d_z (\bm k) = m_1 (k_x^2-k_y^2) + 2 m_2 k_x k_y,
\label{suppl-eq:ConstraintsSpin}
\end{align}
where $a_{0,1}$, $b_{1,2}$ and $m_{1,2}$ are all real. The in-plane components of ${\bm d}$ are odd in ${\bm k}$ and describe the antisymmetric spin-orbit coupling of the Rashba type. This is not surprising since our system is not invariant under the full spatial inversion $I$. The $d_z$ component is even in ${\bm k}$, which is possible in the absence of TR symmetry. 

The continuum Hamiltonian given by Eqs. (\ref{suppl-eq:Hamiltonian}) and (\ref{suppl-eq:ConstraintsSpin}) is applicable in the vicinity of the $\Gamma$ point and indeed in the neighbourhood of any other momentum that is invariant under $C_4K$. On the square lattice, the $M$ point is the only other $C_4K$-invariant momentum.

\subsection{Tight-binding model} 

\begin{figure}
\centering
\begin{tabular}{ccc}
\begin{overpic}[width=0.26\linewidth]{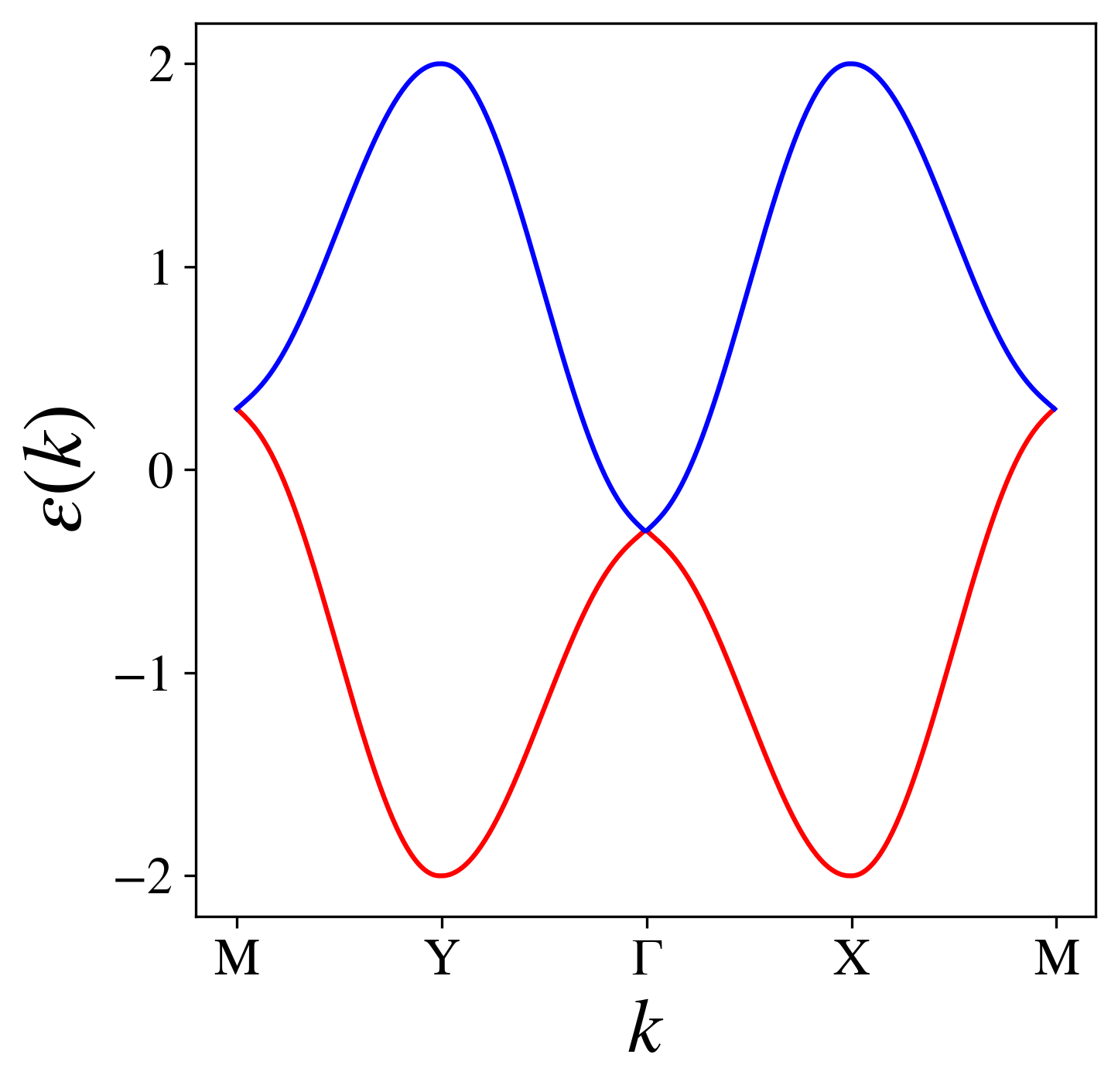}
\put(0,100){\rm{(a)}} 
\end{overpic} & 
\begin{overpic}[width=0.284\linewidth]{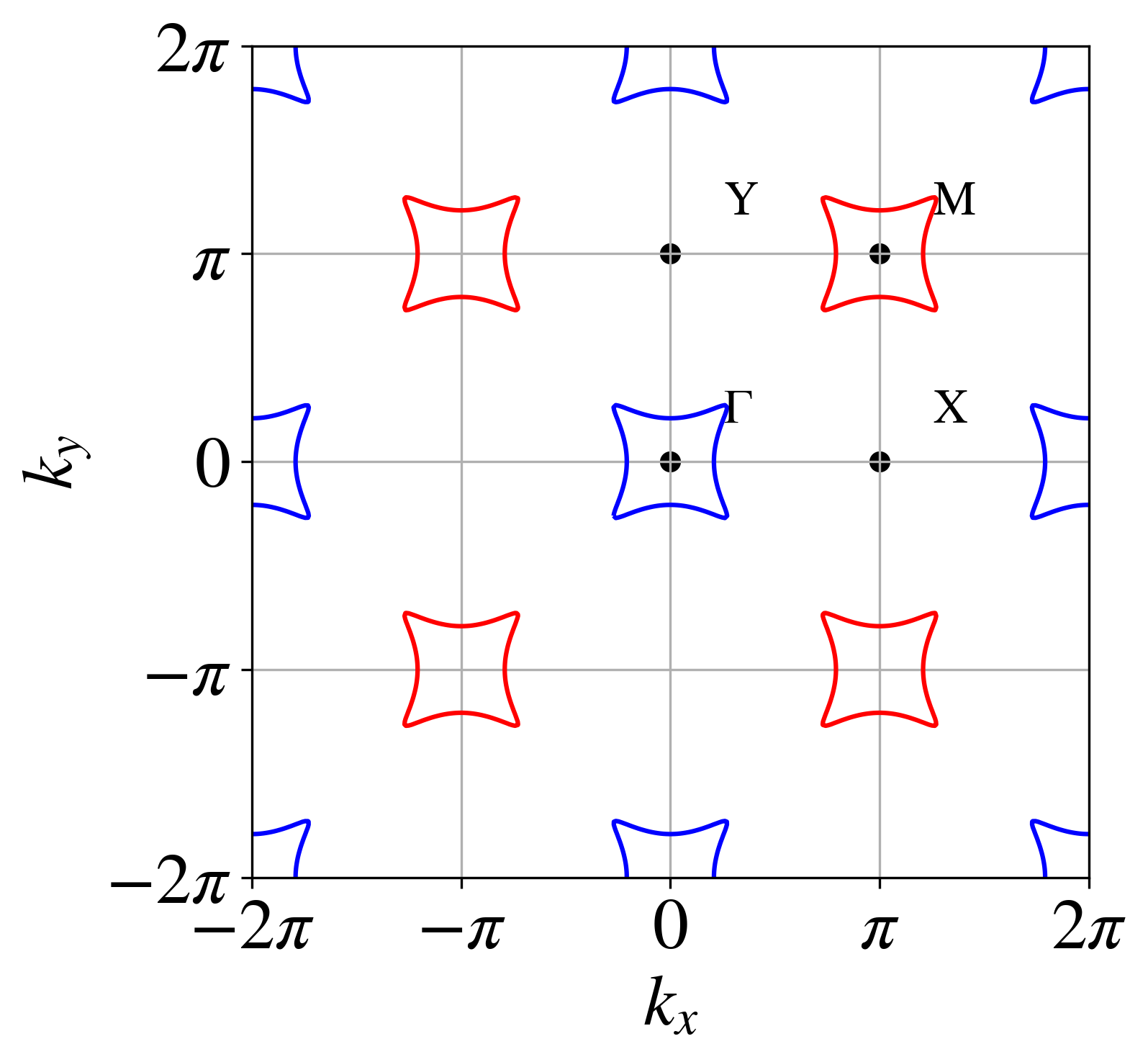}
\put(0,92){\rm{(b)}} 
\end{overpic} &
\begin{overpic}[width=0.284\linewidth]{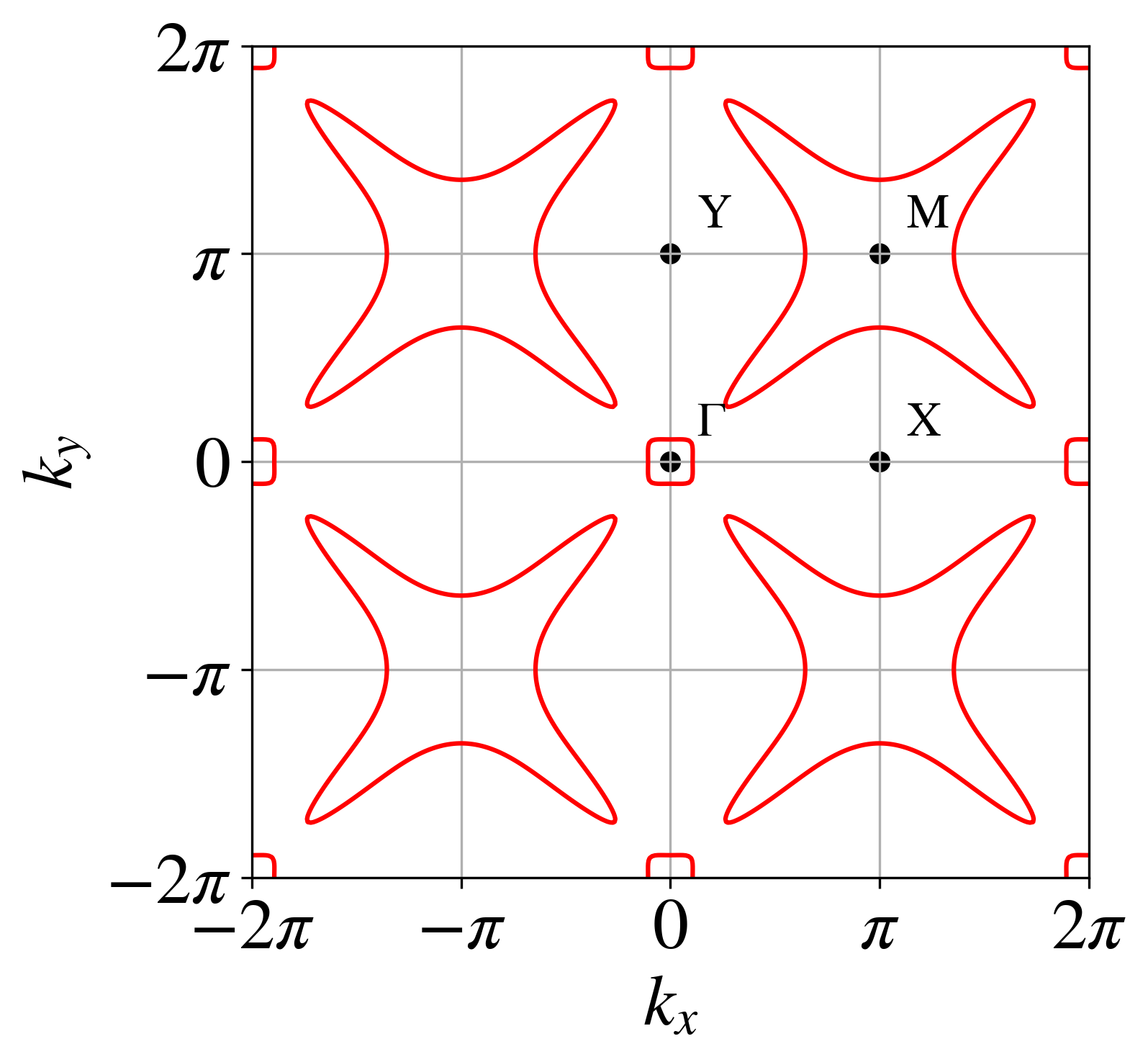}
\put(0,92){\rm{(c)}}  
\end{overpic} 
\end{tabular}
\caption{(a) Band structure for $\lambda=0.4$, $J_1 =1, J_2= 0$, and $t=0.15$. Fermi pockets for (b) $\mu =0$ and (c) $\mu =-0.4$.} 
\label{fig1}
\end{figure}

\begin{figure}
\centering
\begin{tabular}{ccc}
\begin{overpic} 
[width=0.26\linewidth]{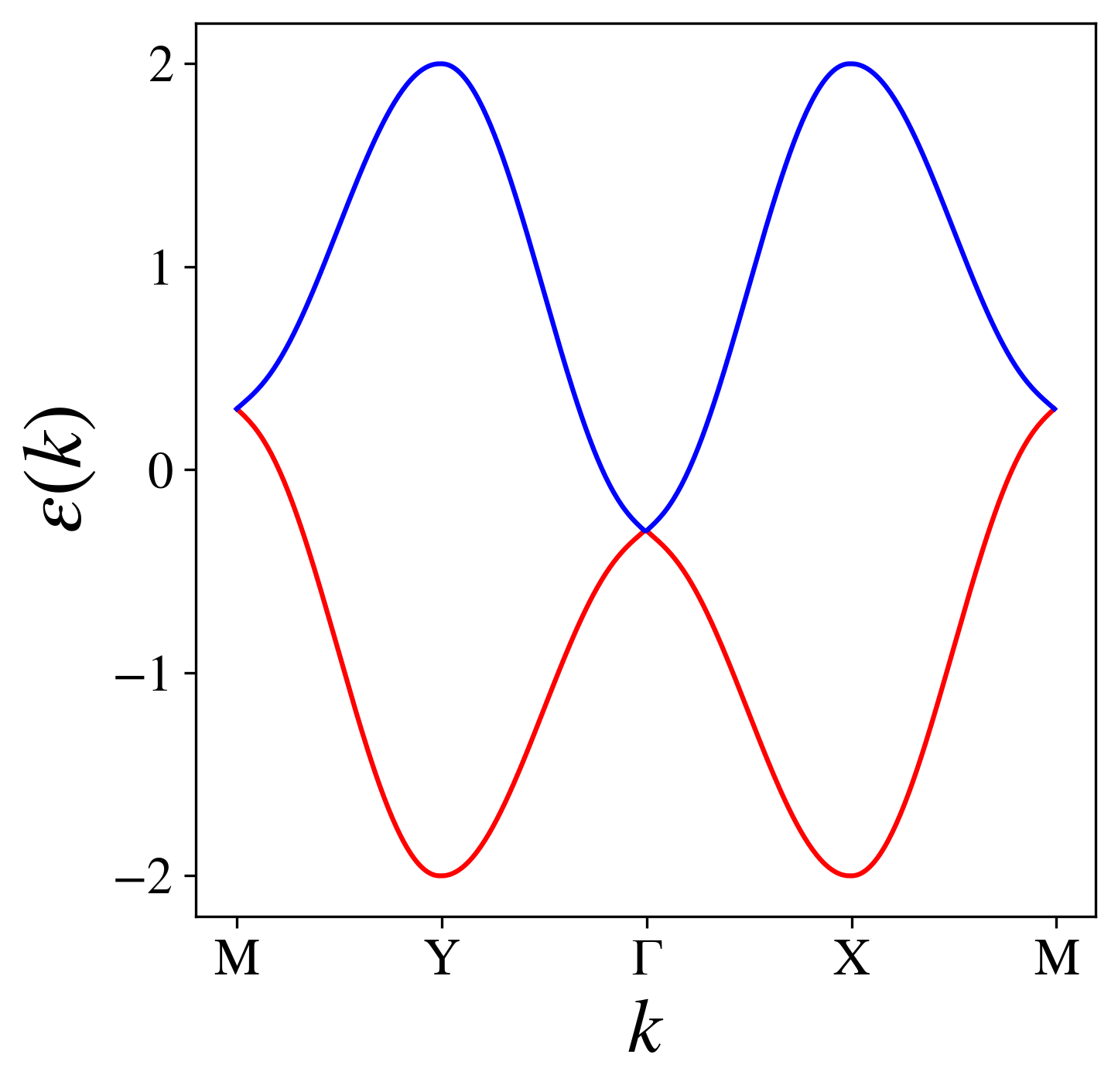}
\put(0,100){\rm{(a)}} 
\end{overpic} & 
\begin{overpic}
[width=0.284\linewidth]{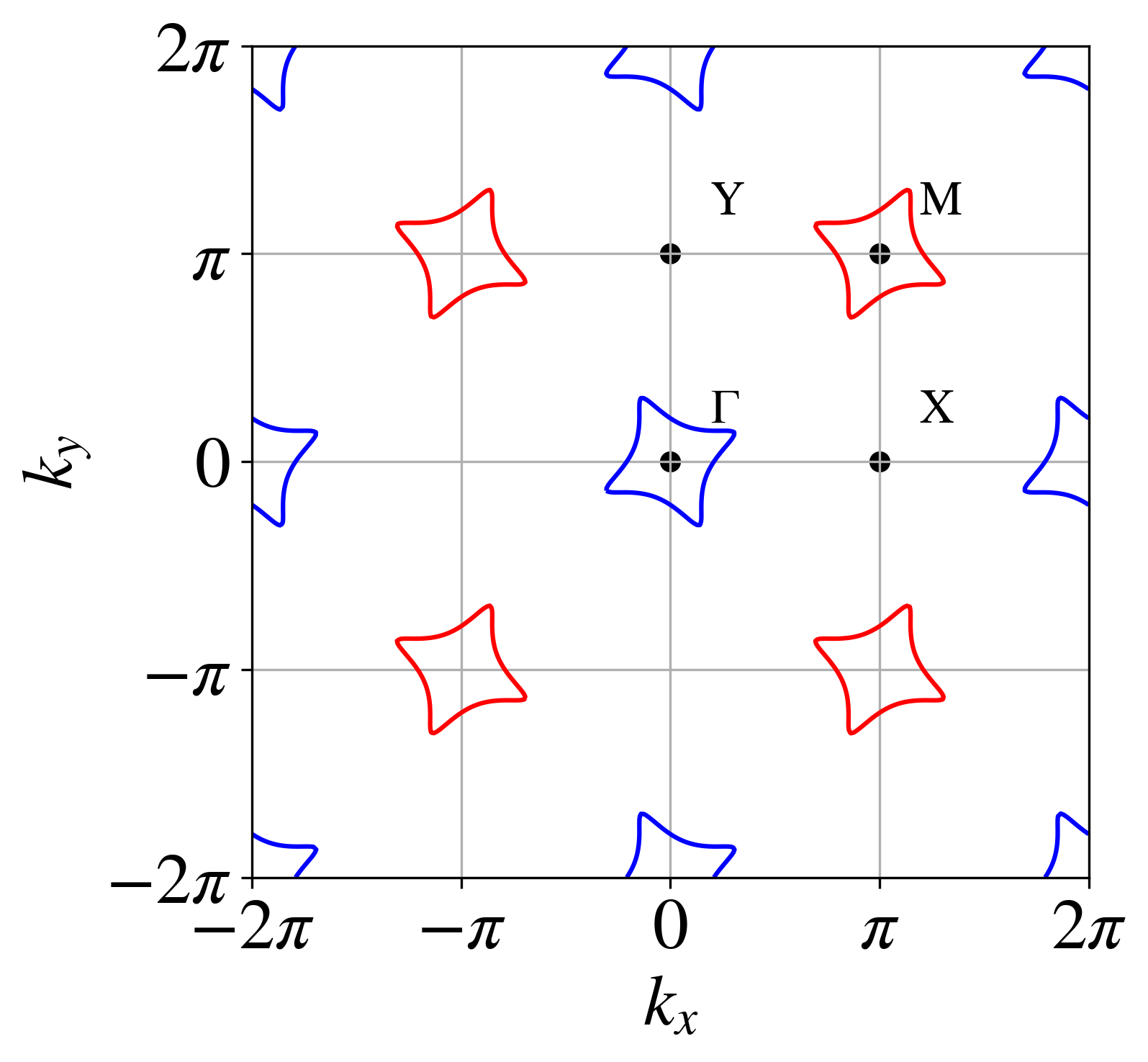}
\put(0,92){\rm{(b)}} 
\end{overpic} &
\begin{overpic}
[width=0.284\linewidth]{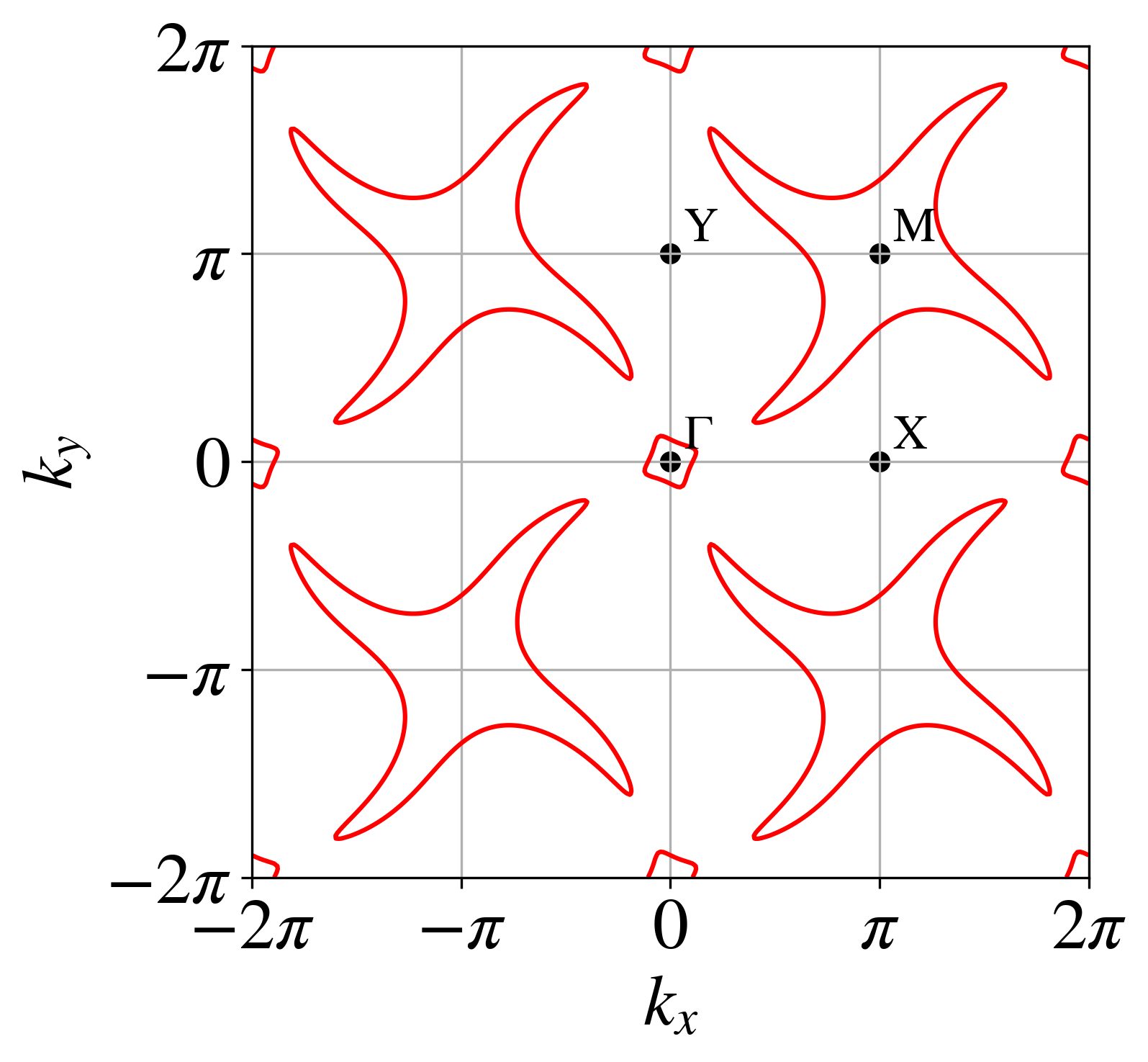}
\put(0,91){\rm{(c)}}  
\end{overpic} 
\end{tabular}
\caption{(a) Band structure for $\lambda=0.4$, $J_1 =J_2= 1$, and $t=0.15$. Fermi pockets for (b) $\mu =0$ and (c) $\mu =-0.4$.} 
\label{fig2}
\end{figure}

To better understand the origin of the various terms in the Hamiltonian, we construct a tight-binding model that is consistent with the $C_4K$ symmetry. We start with a 2D square lattice, as shown in Fig. 2 in the main text. Apart from the usual hopping processes, we have spin-dependent hopping between next-nearest neighbours. These could arise from the Rashba spin-orbit coupling. We introduce ``altermagnetic order parameters'', $J_1$ and $J_2$. They encode preferential hopping of each spin along nearest and next-nearest neighbour bonds. Crucially, the $J_1$ and $J_2$ processes break $C_4$ and $K$, but preserve $C_4 K$. The Hamiltonian is
\begin{align}
    H =& -\frac{t}{2}\sum_{i,\alpha}\big(c^\dagger_{i,\alpha} c_{i+x, \alpha} +c^\dagger_{i,\alpha} c_{i+y, \alpha} + H. c.  \big)  \nonumber\\
    &-i \frac{\lambda}{2} \sum_{i,\alpha \beta}(c^\dagger_{i,\alpha} {\sigma}_{\alpha\beta}^y c_{i-x+y,\beta}-c^\dagger_{i-x+y,\beta} {\sigma}^{y}_{\alpha\beta} c_{i,\alpha})  \nonumber\\ 
     &-i \frac{\lambda}{2}\sum_{i,\alpha \beta} (c^\dagger_{i,\alpha} {\sigma}_{\alpha\beta}^x  c_{i+x+y,\beta}-c^\dagger_{i+x+y,\beta} {\sigma}^{x}_{\alpha\beta} c_{i,\alpha}) \nonumber\\
    &+ \frac{J_1}{2} \sum_{i,\alpha}\big(c^\dagger_{i,\alpha} {\sigma}^z_{\alpha \beta} c_{i+x, \beta} -c^\dagger_{i,\alpha} {\sigma}^z_{\alpha \beta} c_{i+y, \beta} + H. c. \big) \nonumber\\ 
&+\frac{J_2}{2} \sum_{i,\alpha}\big(c^\dagger_{i,\alpha} {\sigma}^z_{\alpha \beta} c_{i-x+y, \beta} -c^\dagger_{i,\alpha} {\sigma}^z_{\alpha \beta} c_{i+x+y, \beta} + H. c.  \big),
\end{align}
where $c^\dagger_{i,\alpha}$ creates a particle with spin $\alpha$ at lattice site $i$, $t$ denotes the hopping parameter, and $\lambda$ corresponds to the spin-orbit coupling. In momentum space, this Hamiltonian takes the form 
\begin{align}
    \hat H(\bm k) =& -t \big(\cos    k_x + \cos    k_y \big) \hat\sigma_0 +\frac{\lambda}{2} \left[\sin    \left(k_x+k_y\right)\hat\sigma_x + \sin    \left(k_y-k_x\right)\hat\sigma_y \right] \nonumber\\ 
  &+ \left[ J_1(\cos    k_x - \cos    k_y) + J_2 \sin    k_x\sin    k_y \right]\hat\sigma_z.
  \label{Eq:HamiltonianMomentum} 
\end{align}
Upon diagonalizing this Hamiltonian, the energy eigenvalues are obtained as 
\begin{align}
&\epsilon_\pm({\bm k}) =\delta \pm \frac{1}{2} \sqrt{\xi + \beta+\gamma},
\label{eq.eps}
\end{align}
where
\begin{align}
&\delta = -t (\cos    k_x + \cos    k_y), \nonumber\\
&   \xi = 4 J_1^2 + J_2^2 + \lambda^2 - 8  J_1^2 \cos    k_x \cos    k_y, \nonumber\\ 
&\beta = (2 J_1^2 - J_2^2) (\cos2     k_x + \cos 2    k_y) + (J_2^2 - \lambda^2) \cos2    k_x \cos2    k_y, \nonumber\\ 
&\gamma =  4 J_1 J_2 (\sin2    k_x \sin    k_y -  \sin    k_x \sin 2    k_y).\nonumber
\end{align}
Here $\epsilon_\pm({\bm k})$ denote the energy of upper and lower bands. The energy bands along the $M-Y-\Gamma-X-M$ are illustrated in Figs.~\ref{fig1}-\ref{fig2}. The two high-symmetry points of Eq.~(\ref{Eq:HamiltonianMomentum}), $\Gamma=(0,0)$ and $M=(\pi,\pi)$ host gapless Dirac nodes with $\epsilon_\Gamma =-2t$ and $\epsilon_M = 2t$. When $t$ is small and the chemical potential is close to zero, we obtain two Fermi pockets around each Dirac point, see Figs.~\ref{fig1}-\ref{fig2}. 

To connect with the long-wavelength form of Eq.~(\ref{suppl-eq:ConstraintsSpin}), we examine the form of the tight-binding Hamiltonian near the Dirac nodes. Near $\Gamma$, keeping terms up to second order in ${\bm k}$, the Hamiltonian (\ref{Eq:HamiltonianMomentum}) takes the form 
\begin{align}
    H_{\Gamma}({\bm k}) =& -\frac{t  }{2}(4- k^2) \hat{\sigma}_0 + \frac{\lambda  }{2} (k_x+k_y) \hat{\sigma}_x + \frac{\lambda    }{2} (k_y-k_x) \hat{\sigma}_y -\left[\frac{J_1   }{2} (k^2_x-k^2_y) - J_2    k_x k_y \right] \hat{\sigma}_z, 
\label{SupplEq:H-Gamma}
\end{align}
whereas near the $M$ point we have
\begin{align} 
    H_M(\bm k) = \frac{t   }{2}(4- k^2)\hat{\sigma}_0 + \frac{\lambda    }{2} (k_x+k_y) \hat{\sigma}_x +\frac{\lambda   }{2}(k_y-k_x)\hat{\sigma}_y+ \left[\frac{J_1   }{2}(k_x^2-k_y^2) + J_2    k_xk_y \right] \hat{\sigma}_z. 
\label{SupplEq:H-M}
\end{align} 
Near the two Dirac points, the band dispersions of Eq.~(\ref{eq.eps}) take the form 
\begin{align}
    \epsilon_{\pm}^\Gamma(\bm k) =  -\frac{t}{2}(4 -k^2) \pm \frac{1}{2} \sqrt{ \big[J_1 ( k_x^2 - k_y^2 ) - 2J_2 k_x k_y\big]^2 + 2\lambda^2 k^2}
\label{SupplEq:epsilon-pm-Gamma}
\end{align}
and 
\begin{align}
    \epsilon_{\pm}^{\rm M}(\bm k) =  \frac{t}{2}(4 -k^2) \pm \frac{1}{2} \sqrt{ \big[J_1 ( k_x^2 - k_y^2 ) + 2 J_2 k_x k_y\big]^2 + 2\lambda^2 k^2}.
\label{SupplEq:epsilon-pm-M}
\end{align}

\subsection{Berry curvature}
In a two-level system, the Hamiltonian can be represented as $\hat H = d_0({\bm k})\hat\sigma_0 + {\bm d}(\bm k)\cdot\hat{\bm\sigma}$. The associated Berry curvature is given by~\cite{BerryReview} 
\begin{align}
    \Omega_{ij}^{\pm} =\pm\frac{ 1}{2|{\bm d}(\bm k)|^3} {\bm d}(\bm k) \cdot \left[\frac{\partial{\bm d}(\bm k)}{\partial k_i} \times \frac{\partial{\bm d}(\bm k)}{\partial k_j} \right],
\label{Eq:BerryC}
\end{align}
where $\pm$ indicates the upper and lower bands. In the tight-binding model described above, the Berry curvature at an arbitrary point in the Brillouin zone comes out to be
\begin{align}
    \Omega_{xy}^{\pm} = \mp\frac{\lambda^2  (\cos    k_x + \cos    k_y) [ J_1  (\cos    k_x- \cos    k_y) + J_2  \sin    k_x \sin    k_y ]}{8\left(\lambda^2 \big(1- \cos    k_x \cos    k_y \big) + [J_1  (\cos    k_x - \cos    k_y)  - J_2 \sin    k_x \sin    k_y]^2\right)^{3/2}}.
    \label{Eq:BerryLattice+}
\end{align}

\begin{figure}
\centering
\begin{tabular}{ccc}
\begin{overpic}
[width=0.31\linewidth]{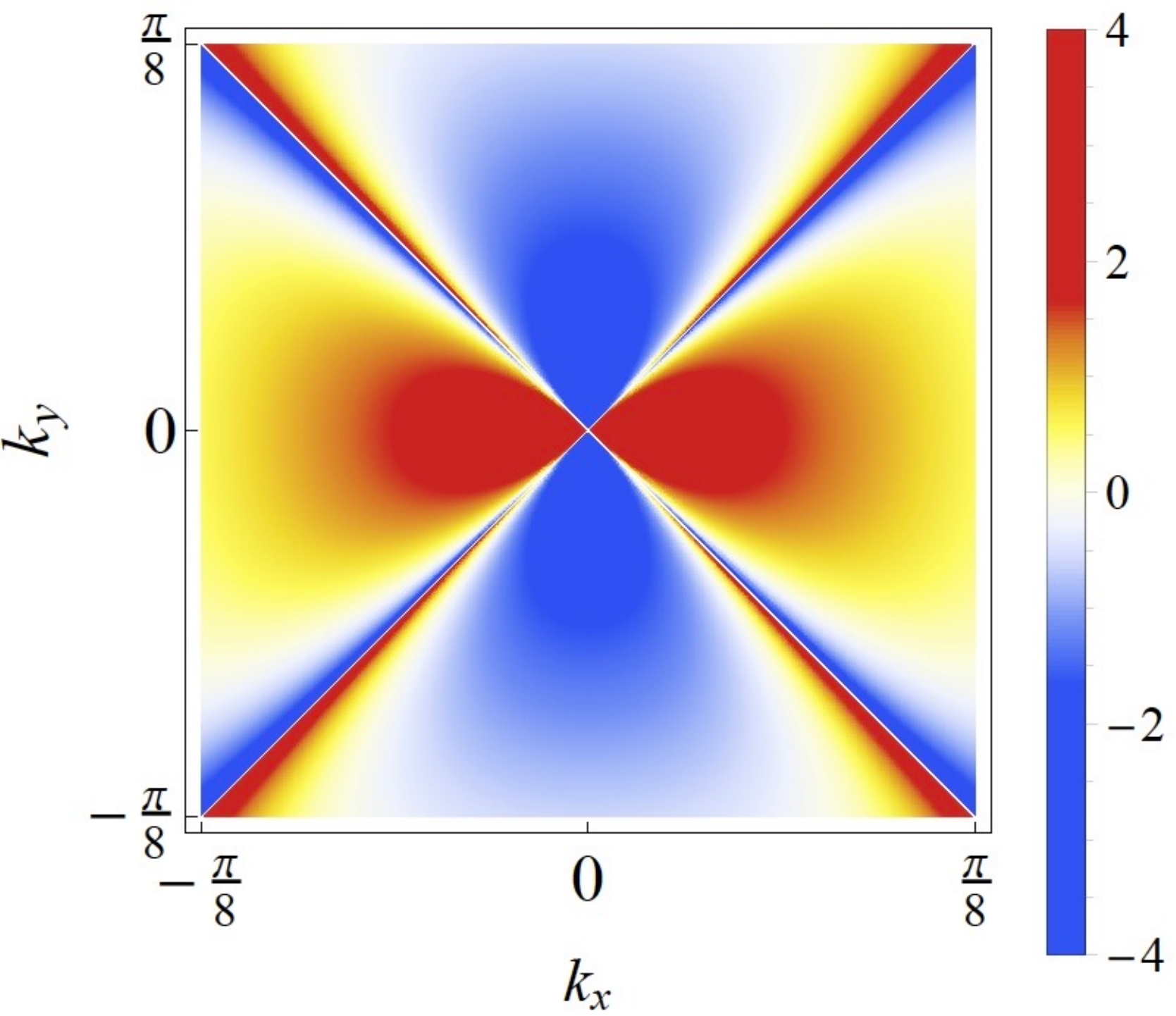}
\put(0,92){\rm{(a)}} 
\end{overpic} & 
\begin{overpic}
[width=0.31\linewidth]{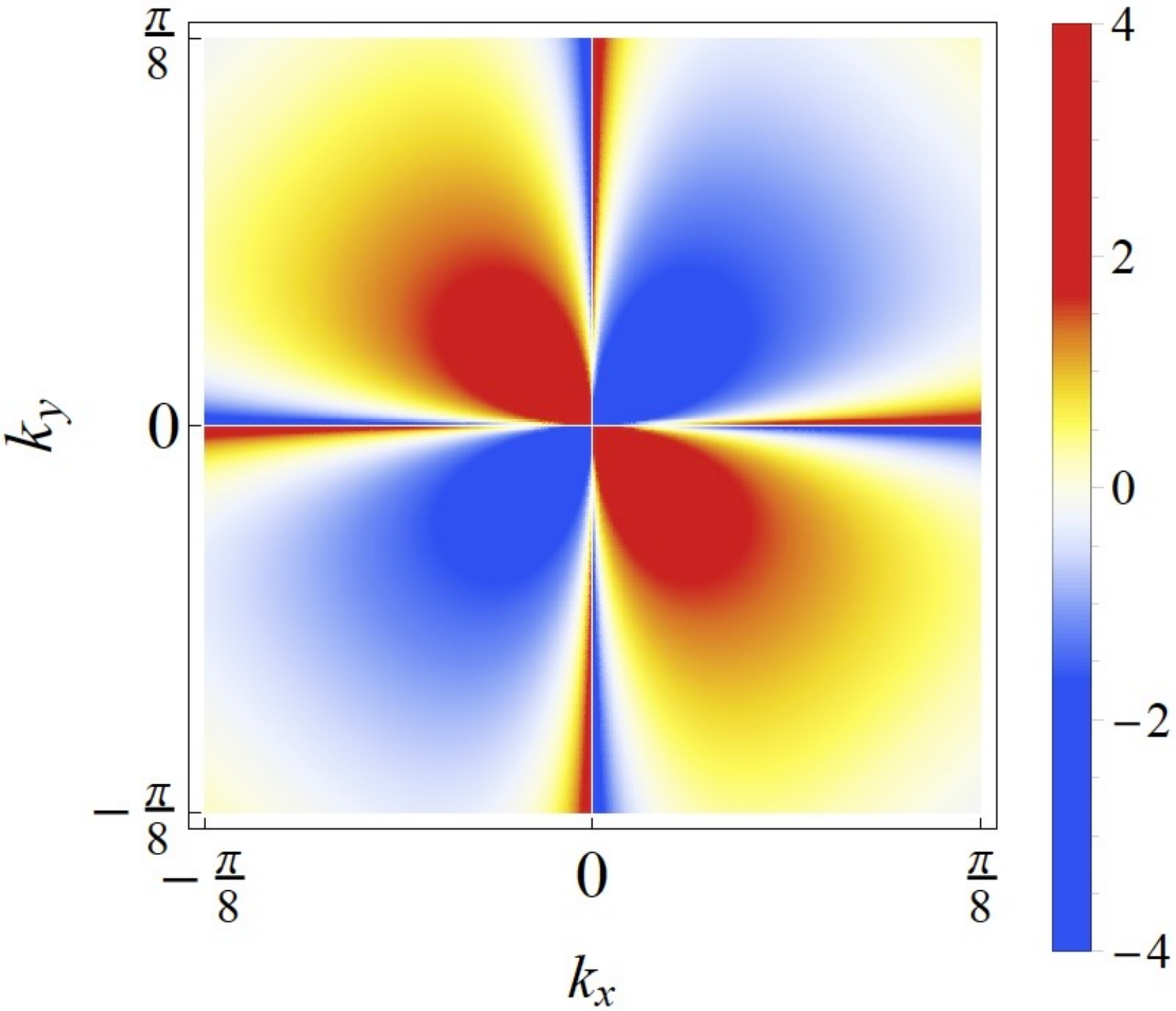}
\put(0,92){\rm{(b)}} 
\end{overpic}
 & 
\begin{overpic}
[width=0.31\linewidth]{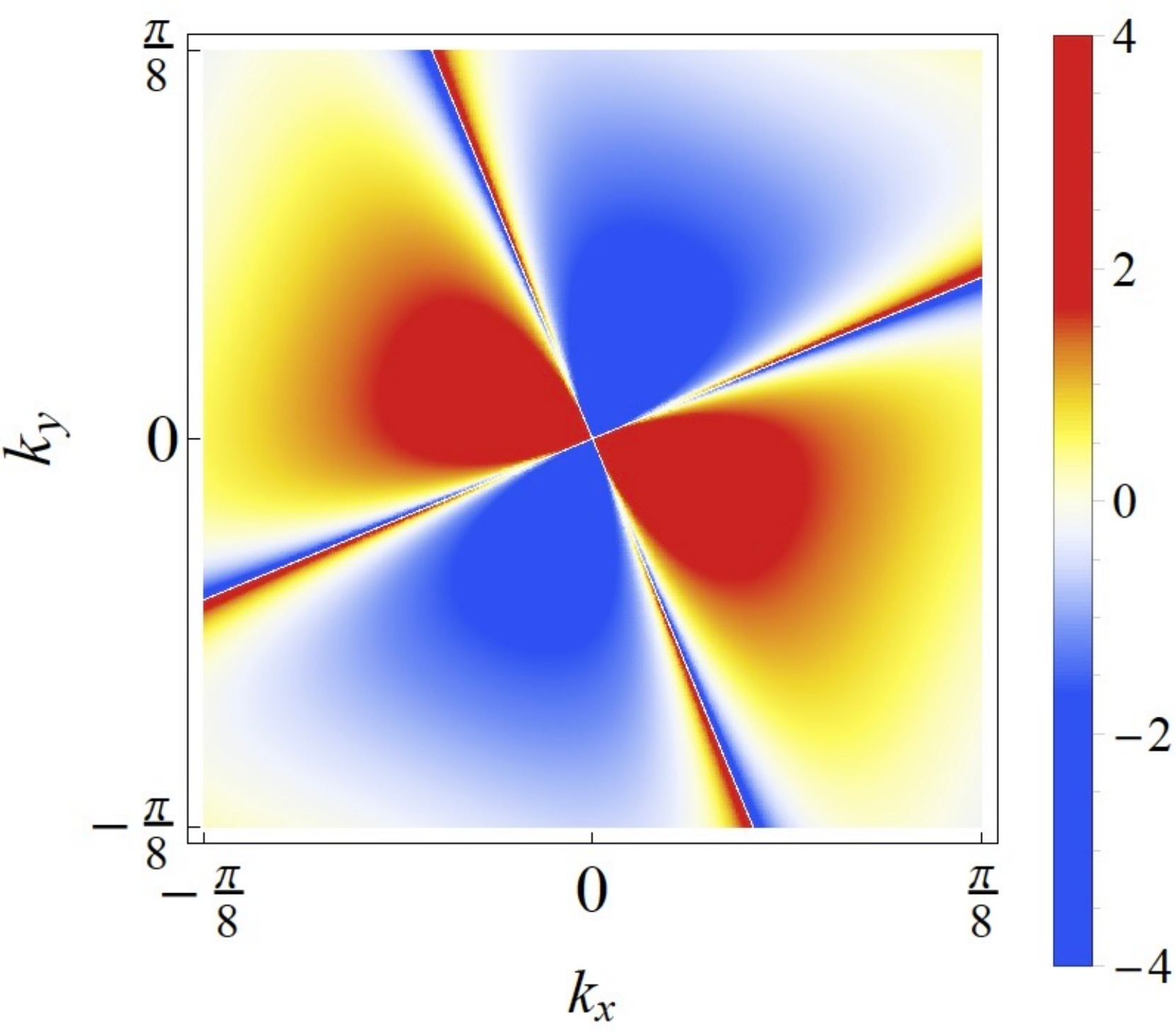}
\put(0,92){\rm{(c)}} 
\end{overpic}
\end{tabular}
\caption{Quadrupole-like distribution of the Berry curvature in the upper band around $\bm{k}=0$, for (a)  $J_1 \neq 0$ and $J_2=0$, (b) $J_1=0$ and $J_2 \neq 0 $ and (c) $J_1 = J_2 \neq 0 $. The $C_4K$ symmetry is clearly seen. The distributions in (a) and (b) show $d_{x^2-y^2}$ and $d_{xy}$ character, respectively.
}
\label{fig:Berryquad}
\end{figure}

We immediately find $\Omega_{xy}^+ = - \Omega_{xy}^-$. In a $d_{xy}$-type altermagnet ($J_1=0$, $J_2\neq0$), the Berry curvatures in the two Dirac valleys are identical. For the general case of non-zero $J_1$ and $J_2$, the Berry curvatures around the $\Gamma$ and $M$ points come out to be
\begin{align}
   \Omega^{\Gamma}_{\pm} =  \pm \frac{[J_1  (k_x^2 - k_y^2) - 2 J_2    k_x k_y]  \lambda^2   }{\left( [J_1  (k_x^2- k_y^2)-2J_2    k_xk_y]^2  + 2 \lambda^2  k^2 \right)^{3/2}}
\end{align}
and
\begin{align}
   \Omega^{M}_{\pm} =  \mp \frac{[J_1    (k_x^2 - k_y^2)  + 2 J_2    k_x k_y]  \lambda^2   }{\left( [J_1  (k_x^2 - k_y^2)+2J_2   k_xk_y]^2  + 2\lambda^2  k^2 \right)^{3/2}}.
\end{align}
Fig.~\ref{fig:Berryquad} shows $ \Omega^{\Gamma}_{+}$ in the vicinity of the $\Gamma$ point for three representative values of $J_1$ and $J_2$. The plots exhibit clear quadrupole-like distributions.

\section{Spin polarization in the current} 
\label{Sec:SpinPol}

We examine spin polarization in the Bloch states and its impact on the Hall current. 
A two-level Hamiltonian, where the two states correspond to two spin projections along $z$, is given by
$\hat H(\bm k) = d_0(\bm k)\hat{\sigma}_0 + \bm d(\bm k) \cdot \hat{\bm\sigma}$.
For the energy dispersion and the eigenstates we have
\begin{align}
   \epsilon_{\pm}(\bm k) = d_0(\bm k) \pm|\bm d(\bm k)|,\qquad
    \psi_\pm(\bm k) = \left( 
    \begin{matrix}
       u_{\pm,\uparrow}(\bm k) \medskip \\
       u_{\pm,\downarrow}(\bm k)
    \end{matrix}
    \right).
\label{SupplEq:eps-psi-pm}
\end{align}

The electric current can now be considered to have two components, one associated with each spin projection:
\begin{align}
     j_i^\uparrow =-e \int_{\bm k} {\dot{r}}_i |u_{\pm,\uparrow}(\bm k)|^2~f(\bm k), \nonumber\\
     j_i^\downarrow =-e \int_{\bm k} {\dot{r}}_i |u_{\pm,\downarrow}(\bm k)|^2~f(\bm k). \nonumber
     \label{Eq:spin}
\end{align}
From these equations, we obtain the net charge current previously introduced in Eq.~(\ref{Eq:current}) as $j_i^c = j_i^\uparrow + j_i^\downarrow $. We note that $|u_{\pm,\uparrow}(\bm k)|^2+|u_{\pm, \downarrow}(\bm k)|^2 = 1$ due to the normalization of the eigenstates. We now define the spin-polarized component of current as
\begin{align}
    j_{s,i} = j_i^\uparrow - j_i^\downarrow &= -e \int_{\bm k} {\dot{r}}_i \left[|u_{\pm,\uparrow}(\bm k)|^2-|u_{\pm,\downarrow}(\bm k)|^2\right] f(\bm k)\nonumber\\
    &= -e \int_{\bm k} {\dot{r}}_i ~s(\bm k)~ f(\bm k), 
\end{align} 
where we introduced a spin-polarization function in each band as follows:
\begin{equation}
s_\pm(\bm k) = |u_{\pm,\uparrow}(\bm k)|^2 - |u_{\pm,\downarrow}(\bm k)|^2.
\label{SupplEq:s-pm-definition}
\end{equation}
This quantity encodes the expectation value of the spin-$z$ operator in the Bloch state at the wave vector $\bm k$. 

It is convenient to parameterize the vector $\bm d(\bm k)$ by its polar angle $\Theta$ and azimuthal angle $\Phi$, 
\begin{align}
\bm d = |{\bm d}| \left(\sin \Theta \cos \Phi, \sin \Theta \sin \Phi, \cos \Theta \right),
\end{align}
where 
\begin{align}
  &  \sin \Theta = \frac{\sqrt{d_x^2+d_y^2}}{|{\bm d}|},\quad \cos\Theta = \frac{d_z}{|{\bm d}|}, \nonumber\\
    & \sin \Phi = \frac{d_y}{\sqrt{d_x^2+d_y^2}},\quad \cos\Phi = \frac{d_x}{\sqrt{d_x^2+d_y^2}}.\nonumber
\end{align}
Then the eigenstates in Eq. (\ref{SupplEq:eps-psi-pm}) take the form
\begin{align}
\psi_+(\bm k) = \left( 
    \begin{matrix}
       \cos(\frac{\Theta}{2})e^{-i\Phi} \medskip\\
       \sin(\frac{\Theta}{2})
    \end{matrix}
    \right),  \qquad \psi_-(\bm k) = \left( 
    \begin{matrix}
       \sin(\frac{\Theta}{2})e^{-i\Phi} \medskip \\
       -\cos(\frac{\Theta}{2})
    \end{matrix}
    \right).
\end{align}
For the spin-polarization factor (\ref{SupplEq:s-pm-definition}) we have $s_\pm (\bm k) = \pm \cos \Theta$.
Using this last expression with the Hamiltonians (\ref{SupplEq:H-Gamma}) and (\ref{SupplEq:H-M}), we finally obtain
\begin{align}
   s^{ \Gamma}_\pm(\bm k) = \mp \frac{J_1  (k^2_x-k^2_y)-2J_2 k_x k_y }{\sqrt{ [J_1   ( k_x^2 - k_y^2 ) - 2J_2   k_x k_y]^2 + 2\lambda^2  k^2 }}, 
\end{align}
and
\begin{align}
   s^{M}_\pm(\bm k) = \pm \frac{J_1  (k^2_x-k^2_y) +2J_2  k_x k_y }{\sqrt{ [J_1    ( k_x^2 - k_y^2 ) + 2J_2    k_x k_y]^2 + 2\lambda^2  k^2}}
\end{align}
near the Dirac points.

\section{Symmetry-allowed Quadrupole and Drude integrals}

We now consider the two-level system with two Fermi pockets described in the main text and calculate contributions to the charge and spin currents from each Fermi surface. We first note that the system hosts Dirac points at $\Gamma$ and $M$, due to a Kramers-like degeneracy arising from $C_4K$ symmetry. With the Fermi energy close to the Dirac points, we have two small Fermi pockets. In the vicinity of the Dirac points, we denote the momentum in polar coordinates as $(k_x,k_y)=k(\cos\varphi,\sin\varphi)$. The band dispersion of the lower band near the $\Gamma$ point, see Eqs. (\ref{SupplEq:epsilon-pm-Gamma}) and (\ref{SupplEq:epsilon-pm-M}), is given by
\begin{align}
    \epsilon (\bm k) = -2t + \frac{tk^2}{2}
    -\frac{\sqrt{2}}{2}\lambda  k \bigg[1+ \frac{1}{2} \frac{\tilde J^2(\varphi)k^2}{2\lambda^2}-\frac{1}{8} \frac{\tilde J^4(\varphi)k^4}{4\lambda^4 }  + \dots \bigg],
\end{align}
where $\tilde J(\varphi)=J_1\cos 2\varphi-J_2\sin 2\varphi$. 
The leading term in this expression describes an isotropic (circular) Fermi surface and the anisotropies are of the order of $J_{1,2}^2/\lambda^2$. 

Below, we calculate the current contributions assuming that $J_1,J_2\ll\lambda$, which corresponds to nearly-isotropic Fermi surfaces. In order to evaluate currents to the leading order in $J_{1,2}/\lambda$, we need 
\begin{align}
    s(\bm k) \simeq \frac{\tilde J(\varphi)k}{\sqrt{2}\lambda }\bigg[1-\frac{1}{2} \frac{\tilde J^2(\varphi)k^2}{2\lambda^2  }  + \frac{3}{8} \frac{\tilde J^4(\varphi)k^4}{4\lambda^4} \bigg]
\end{align}
and
\begin{align}
    \Omega \simeq -  \frac{\tilde J(\varphi)}{2 \sqrt{2} \lambda   k} \bigg[1-\frac{3}{2}  \frac{\tilde J^2(\varphi)k^2}{2\lambda^2 }  + \frac{15}{8} \frac{\tilde J^4(\varphi)k^4}{4\lambda^4 }\bigg]. 
\end{align} 
The current contributions involve several integrals over the Brillouin zone ($Q_{ij}^c$, $Q_{ij}^s$, $\mathcal{M}_{ijk}^c$ and $\mathcal{M}_{ijk}^s$, see Sec.~\ref{sec.symmconstraints} above). We evaluate these integrals at zero temperature, where the equilibrium distribution function $f_0$ is the step-function in energy. After integration by parts, we may write $\partial_i f_0=-\delta[\epsilon(\bm k)-\mu]\partial_i \epsilon(\bm k) $, where $\mu$ is the chemical potential. Along these lines, each integral in the current can be written in the following form:
\begin{equation}
    I = \frac{1}{(2\pi)^2} \int_0^\infty k \, dk \int_0^{2\pi} d\varphi \, F(k, \varphi) \left(\frac{1}{|\partial_k \epsilon(k)|}\right)_{k=k_F(\varphi)} \delta[k-k_F(\varphi)],
\label{suppl-eq:I-theta k}
\end{equation}
where $F(k,\varphi)$ is some function in 2D polar coordinates,
\begin{align}
    |\partial_k \epsilon(k)&|_{k=k_F(\varphi)}=t k_F(\varphi)  - \frac{\sqrt{2}}{2}\lambda + \frac{ \tilde J^2(\varphi) k_F^2(\varphi) [-24 \lambda^2 + 5  \tilde J^2(\varphi)k^2_F(\varphi)]}{32 \sqrt{2} \lambda^3}
\end{align}
and the Fermi wave vector is given by
\begin{align}
    k_F(\varphi) \simeq  (\mu + 2 t) \left( -\frac{\sqrt{2}}{\lambda} + \frac{1}{w} \left[ \frac{\tilde J^2(\varphi)}{2\lambda^2  }\left(\frac{\mu+2t}{\lambda}\right)^2 - \frac{1}{2} \frac{\tilde J^4(\varphi)}{4\lambda^4 }\left(\frac{\mu+2t}{\lambda}\right)^4\right] \right).
\end{align}
Here $w= {d \epsilon_0} /dk$ is the isotropic Fermi velocity and $\epsilon_0 =-2t + {tk^2}/{2}
    -\sqrt{2}/{2}~\lambda  k$. 
Evaluating the integral over $k$, Eq.~(\ref{suppl-eq:I-theta k}) takes the form 
\begin{align}
   I = \frac{1}{(2\pi)^2} \int_0^{2\pi} d\varphi~ k_F(\varphi)~F(k_F(\varphi), \varphi) \left(\frac{1}{|\partial_k \epsilon(k)|}\right)_{k=k_F(\varphi)}.
\end{align}

We can now evaluate the contributions to the Hall currents. For brevity, we provide explicit expressions for the $\Gamma$ pocket only, in the leading order in $J_{1,2}/\lambda$. Note that each of the following quantities may take non-zero values on symmetry grounds as shown in Sec.~\ref{sec.symmconstraints} above. We have  
\begin{align}
 & Q_{xx}^c =-Q_{yy}^c = \frac{ J_1 }{32 \pi |\epsilon_\Gamma-\mu|}, ~~~~~ Q_{xy}^c = \frac{ J_2    }{32 \pi |\epsilon_\Gamma-\mu|},\nonumber\\
 & \mathcal{M}^c_{xyy}  =  -\frac{\lambda^2   }{32 \pi |\epsilon_\Gamma-\mu| },~~~~~~~~~~~~\mathcal{M}^s_{xxy}=\mathcal{M}^s_{yyy} = -\frac{J_2   }{16 \pi }, 
\end{align}
\begin{align}
\mathcal{M}^c_{xxy} = -\mathcal{M}^c_{yyy} =  -\frac{3}{64\pi}\frac{J_1 J_2   }{ \lambda^2}  |\epsilon_\Gamma-\mu|,
\end{align}
and
\begin{align}
  Q_{xx}^s = Q_{yy}^s = \frac{1}{128 \pi} \frac{(J_1^2+J_2^2)^2  }{\lambda^6} |\epsilon_{\Gamma} - \mu|^2.
\end{align}
Here $\epsilon_\Gamma$ is the degenerate band energy at the $\Gamma$  point.

\section{Estimation of the current magnitudes}
We have demonstrated a light-induced charge Hall current as a signature of $C_4K$ symmetry breaking. We now estimate the magnitude of the signal and discuss its dependence on tunable parameters. 
We consider a hypothetical material with \(C_4K\) symmetry. The parameters employed are as follows. The light-induced electric field is taken to be \(10^7\,\mathrm{V/m}\) as estimated in Ref.~\onlinecite{Nature2020}. We take the static electric field to be $E_y=10^2$ $\mathrm{V/m}$, a realistic value for Hall measurements in metallic samples. The altermagnetic order parameters are \(J_1 = 0.1\,\mathrm{eV}\) and \(J_2 = 0.08\,\mathrm{eV}\), similar to the values proposed for RuO$_2$~\cite{Agterberg2024}. The strength of spin-orbit coupling is taken to be \(0.16\,\mathrm{eV}\), comparable to that in RuO$_2$~\cite{Occhialini2021}. The sample size is \(1\,\mathrm{mm}\times 1\,\mathrm{mm}\times 34\,\mathrm{nm}\)~\cite{SCIADV2024}. We assume \(\lvert \epsilon_{\Gamma} - \mu \rvert \sim 0.1\,\mathrm{eV}\) and \(\lvert \epsilon_{M} - \mu \rvert \sim 0.125\,\mathrm{eV}\). The relaxation time, which can range from femtoseconds to picoseconds in metals~\cite{MarderBook,Wang2021}, is taken to be \(\tau =10^{-14}\,\mathrm{s}\), giving \(\hbar / \tau \approx 0.066\,\mathrm{eV}\). Although these values do not precisely match any single known material, they are realistic and comparable to those in typical experimental setups. The resulting magnitudes of currents are as follows:  $|I^Q_{c,x}| \sim 3.8\, {\rm \mu A}$, $|I^Q_{s,x}| \sim 3\, {\rm \mu A}$, $|I^D_{c,x}| \sim 0.4 \, {\rm  \mu A}$ and $|I^D_{s,x}| \sim 14 \, {\rm \mu A}$.  

\begin{figure}[t]
    \centering
    \begin{tabular}{ccc} 
    \begin{overpic}[width=0.33\linewidth]{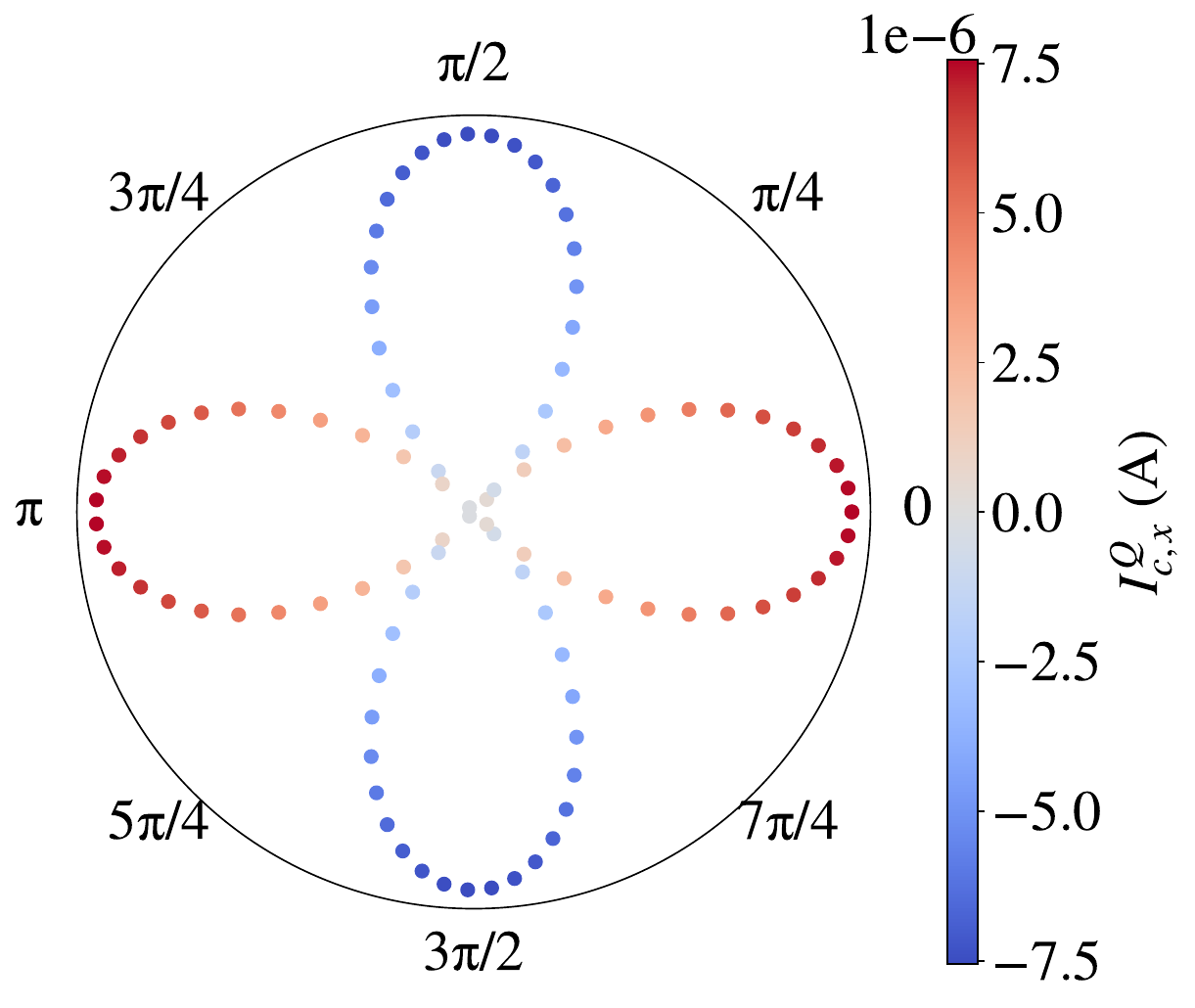}
\put(0,87){\rm{(a)}}
\end{overpic} & 
\begin{overpic}[width=0.33\linewidth]{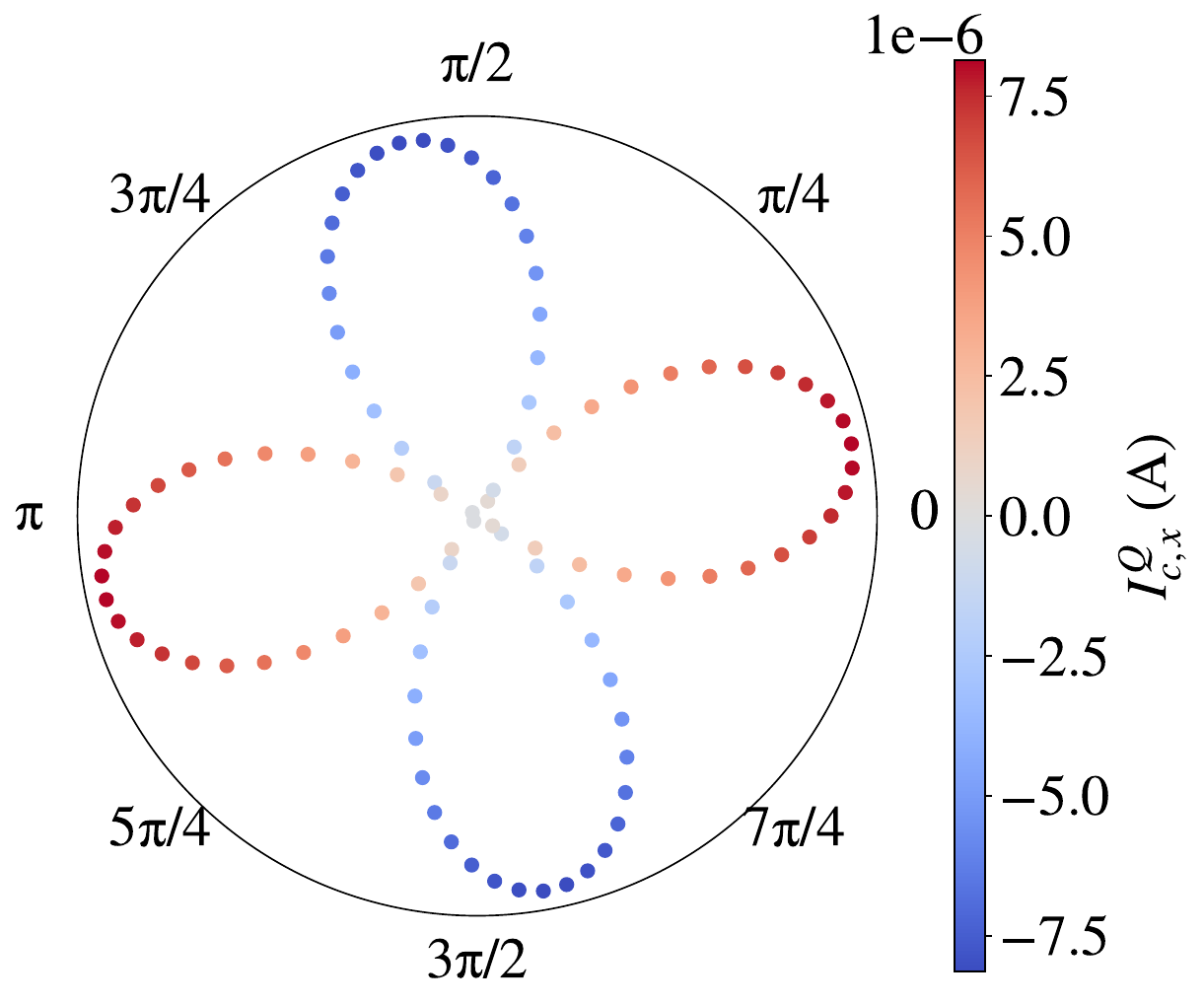}
\put(0,83){\rm{(b)}}
\end{overpic}& 
\begin{overpic}[width=0.33\linewidth]{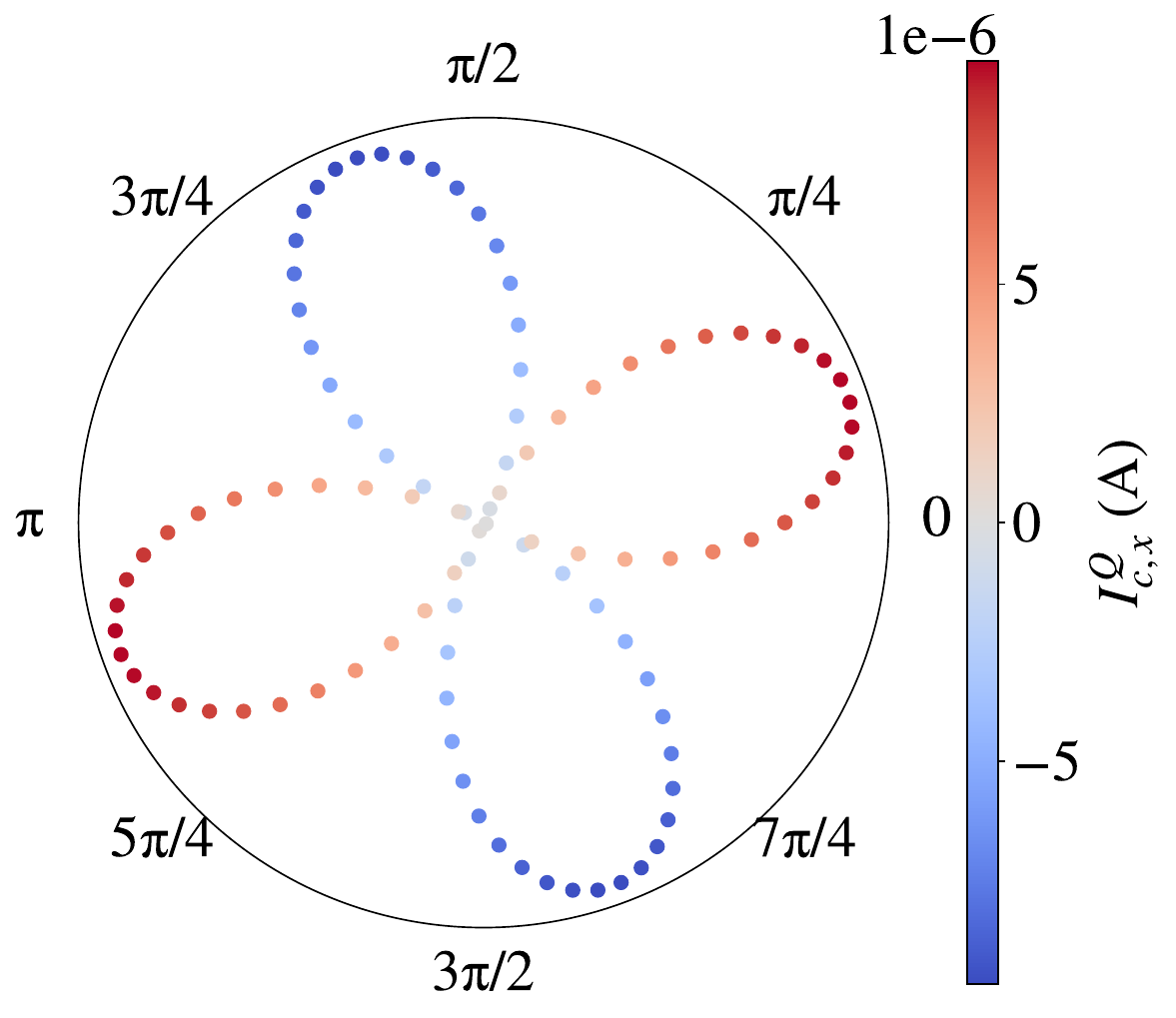}
\put(0,83){\rm{(c)}}
\end{overpic} 
\end{tabular}
\caption{Polar plots of the Berry curvature quadrupole contribution to the charge current vs $\theta$, under different polarization conditions: 
(a) circular polarization (\(\cos \phi = 0\)), 
(b) elliptical polarization (\(\cos \phi = 1/2\)), 
and 
(c) linear polarization (\(\cos \phi = 1\)). The plots are characteristic of $C_4 K$ symmetry, with four lobes that have alternating signs.}   
\label{fig:polarplotQ}
\end{figure}

\begin{figure}[t]
    \centering
    \begin{tabular}{ccc} 
    \begin{overpic}[width=0.33\linewidth]{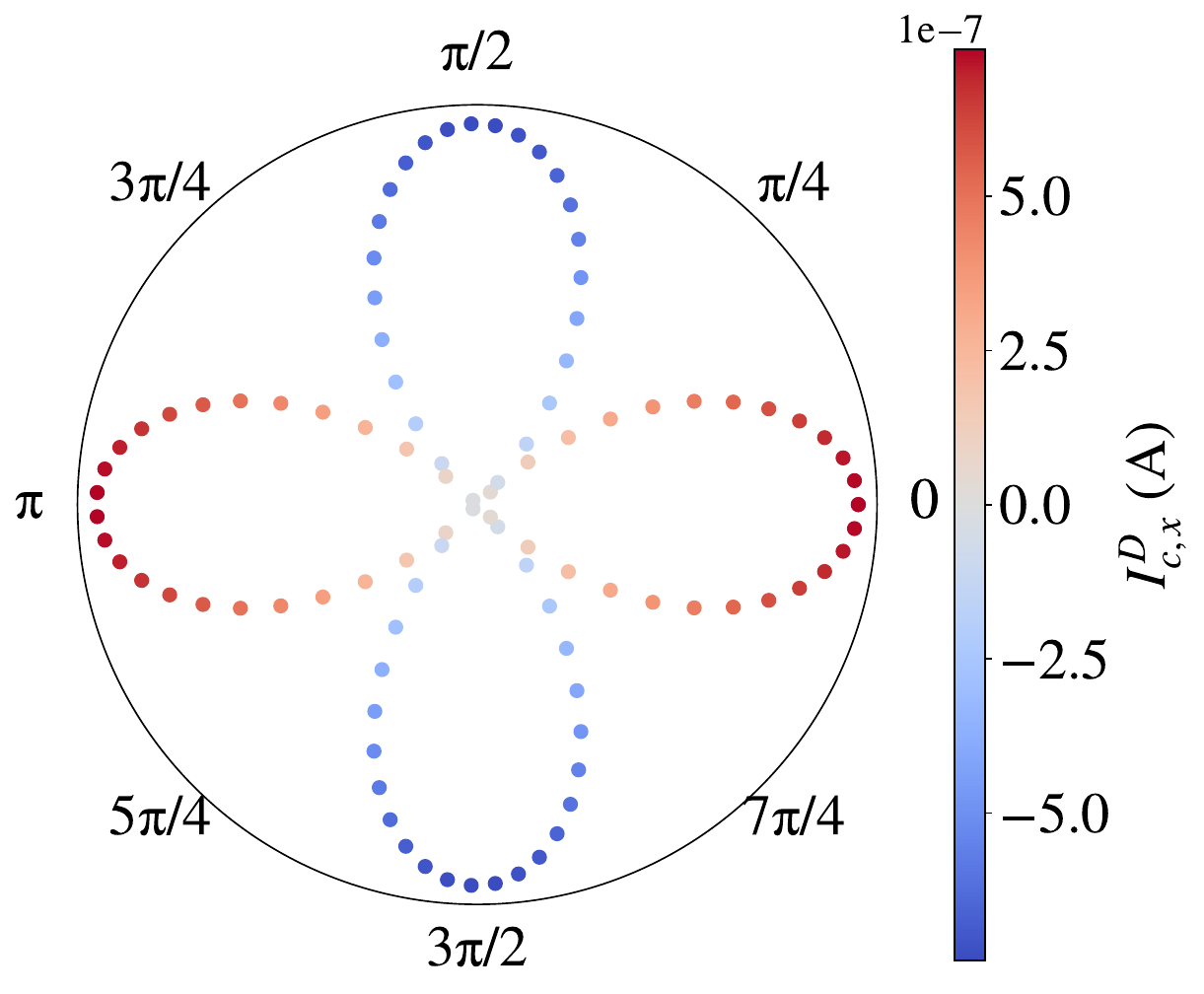}
\put(0,87){\rm{(a)}}
\end{overpic} & 
\begin{overpic}[width=0.33\linewidth]{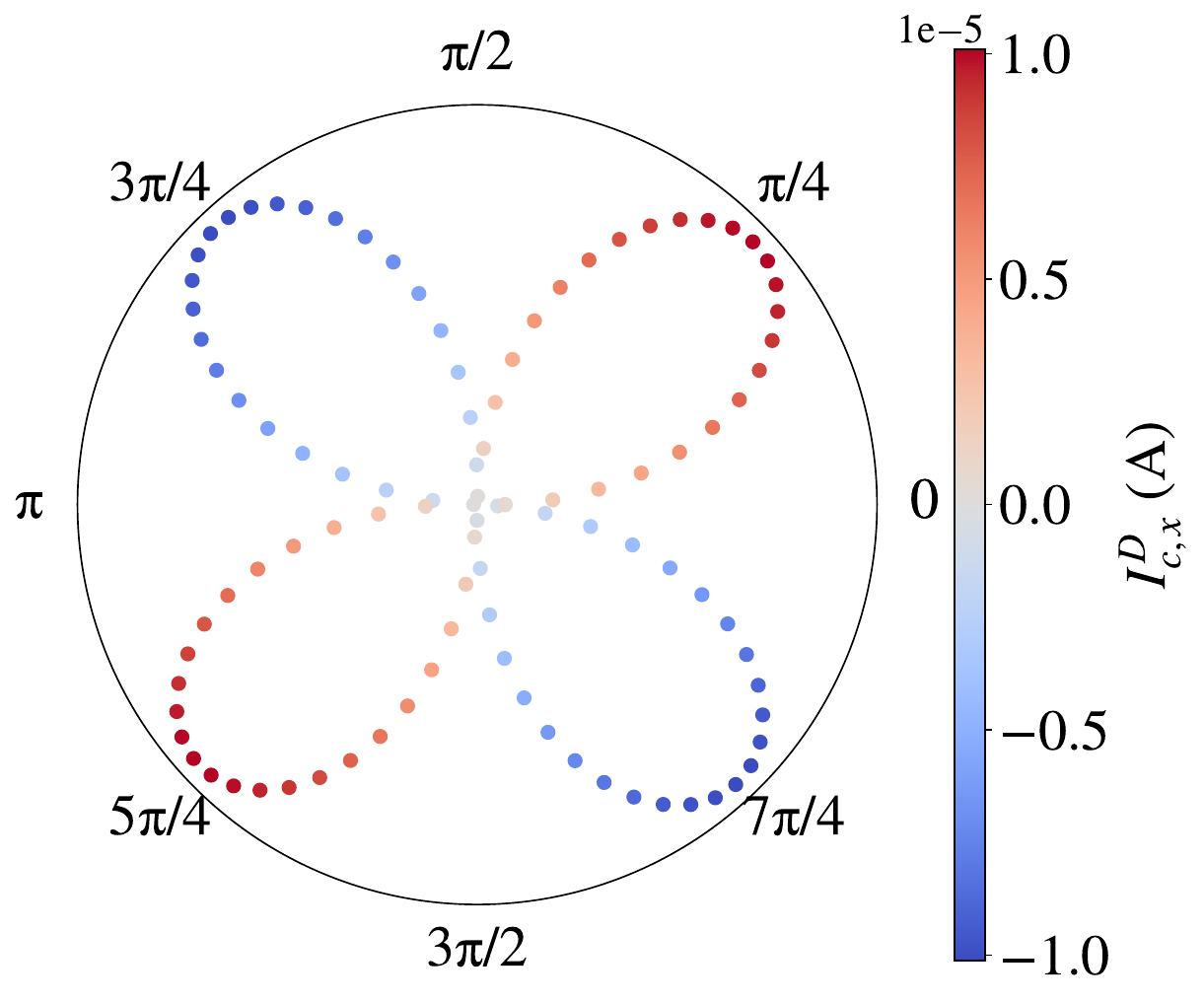}
\put(0,83){\rm{(b)}}
\end{overpic}& 
\begin{overpic}[width=0.33\linewidth]{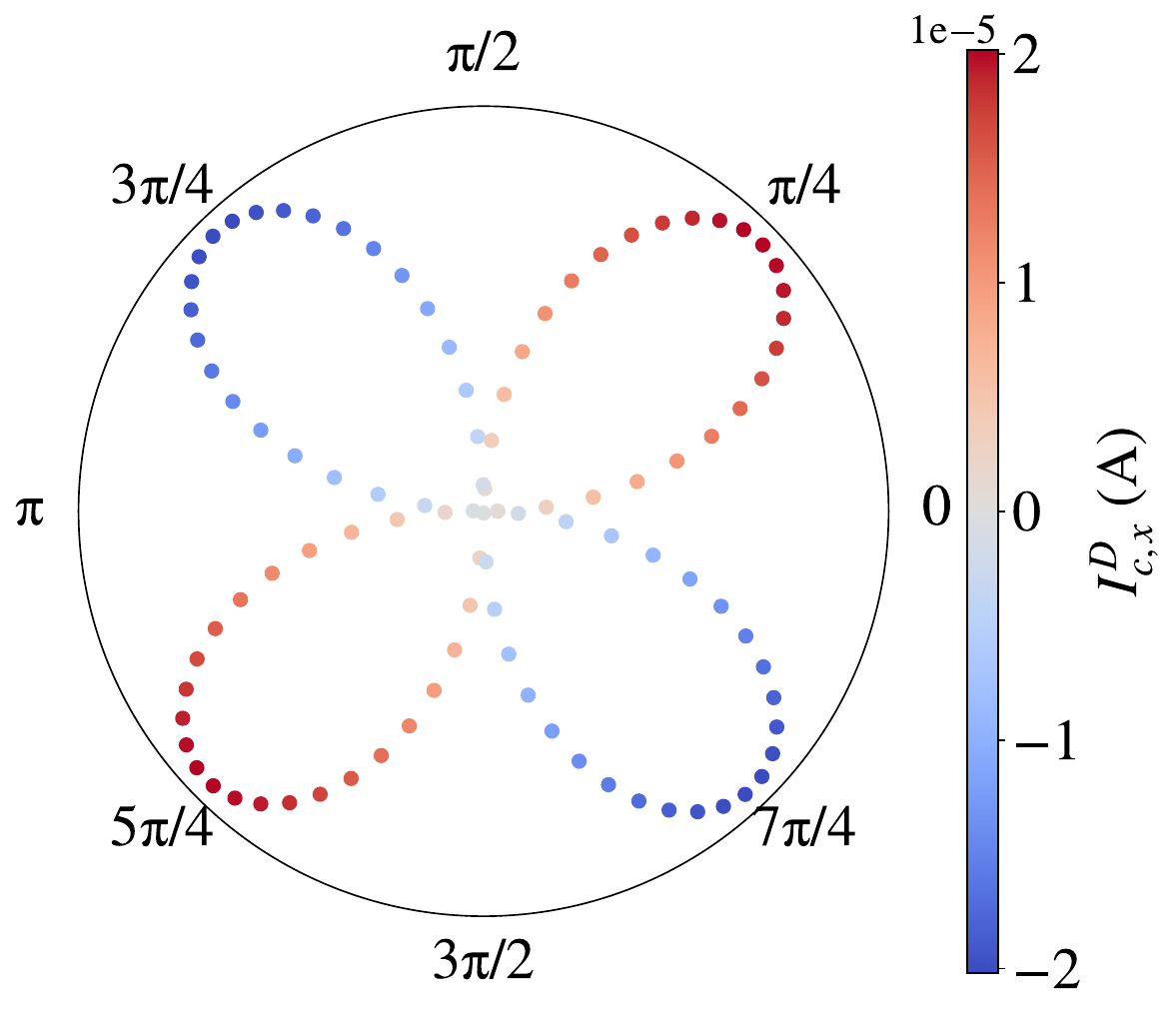}
\put(0,83){\rm{(c)}}
\end{overpic} 
\end{tabular}
\caption{Polar plots of the Drude contribution to the charge current vs $\theta$, under three polarization conditions: 
(a) circular polarization (\(\cos \phi = 0\)), 
(b) elliptical polarization (\(\cos \phi = 1/2\)), 
and 
(c) linear polarization (\(\cos \phi = 1\)).}  
\label{fig:polarplotD}
\end{figure}

\begin{figure}
    \centering
    \begin{tabular}{ccc} 
    \begin{overpic}[width=0.33\linewidth]{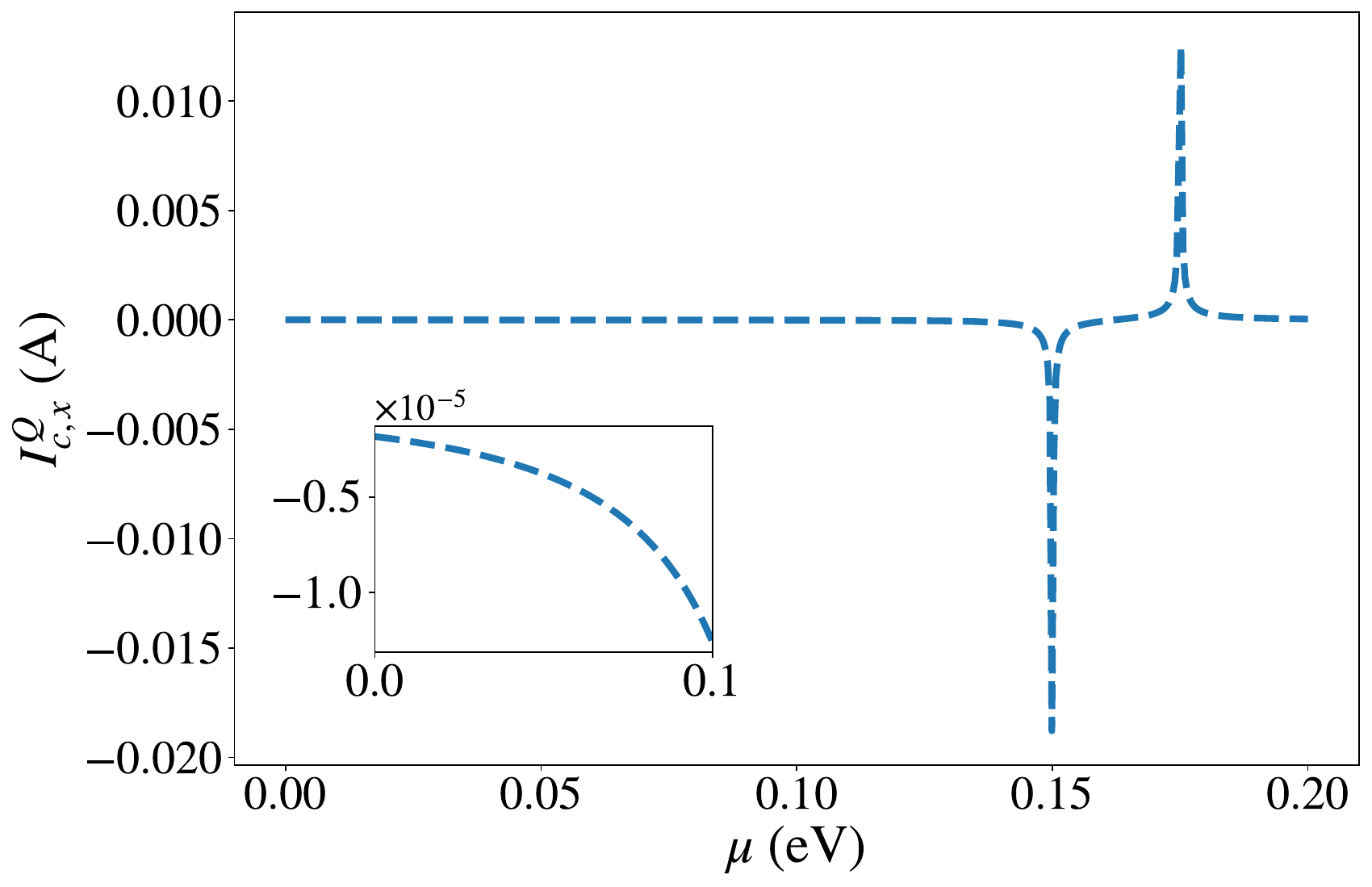}
\put(0,67){\rm{(a)}}
\end{overpic} & 
\begin{overpic}[width=0.33\linewidth]{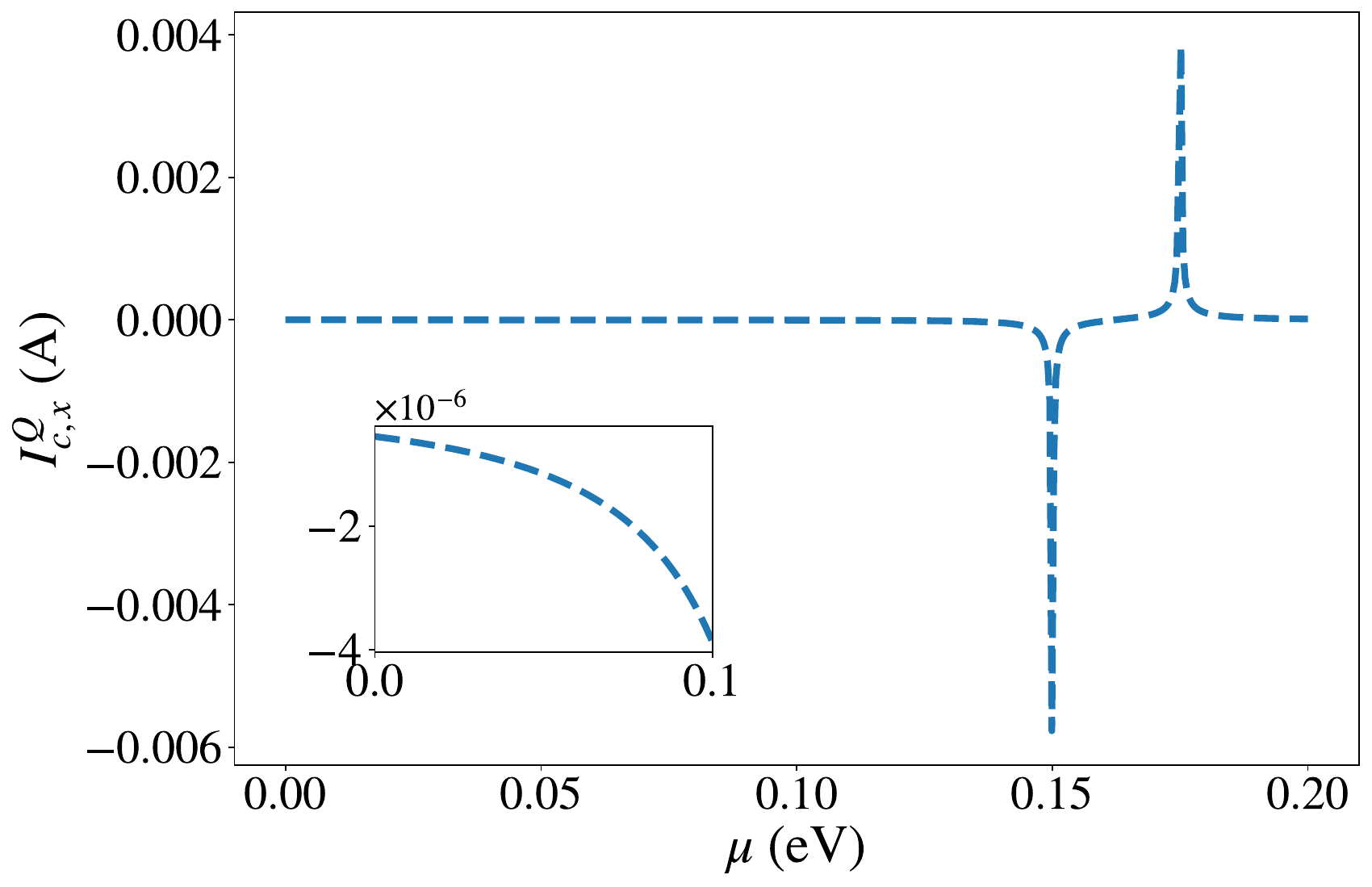}
\put(0,67){\rm{(b)}}
\end{overpic}& 
\begin{overpic}[width=0.33\linewidth]{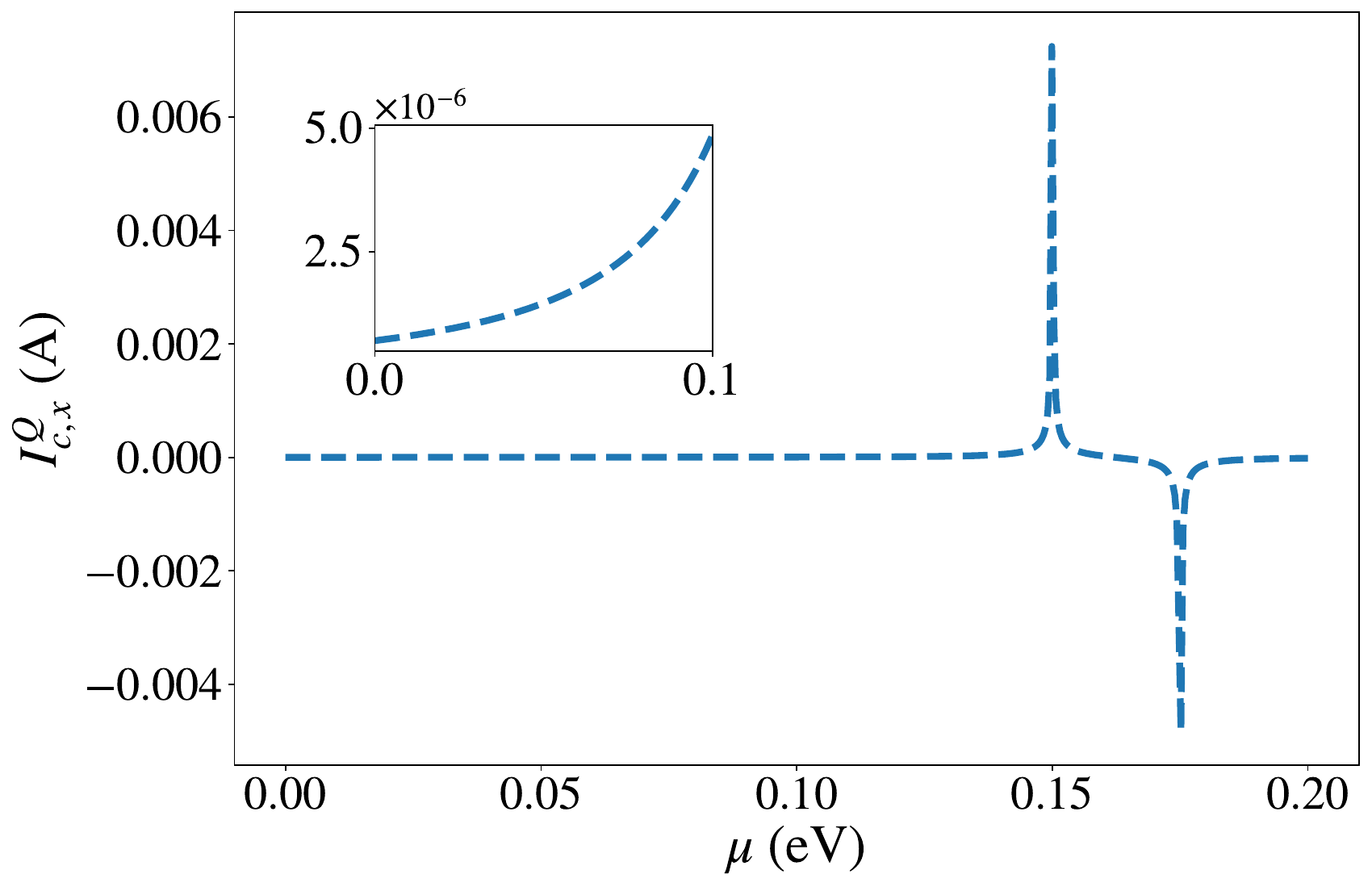}
\put(0,67){\rm{(c)}}
\end{overpic} 
\end{tabular}
\caption{Dependence of the Berry curvature quadrupole contribution to the charge current on the chemical potential. For the purpose of illustration, we fix \(\theta = \pi/3\);
(a) circular polarization (\(\cos \phi = 0\)),
(b) elliptical polarization (\(\cos \phi = 1/2\)),
and 
(c) linear polarization (\(\cos \phi = 1\)). The sharp peaks are where the chemical potential crosses the Dirac points, resulting in a diverging Berry curvature.}  
\label{fig:ChemicalQC}
\end{figure}

\begin{figure}
    \centering
    \begin{tabular}{ccc} 
    \begin{overpic}[width=0.33\linewidth]{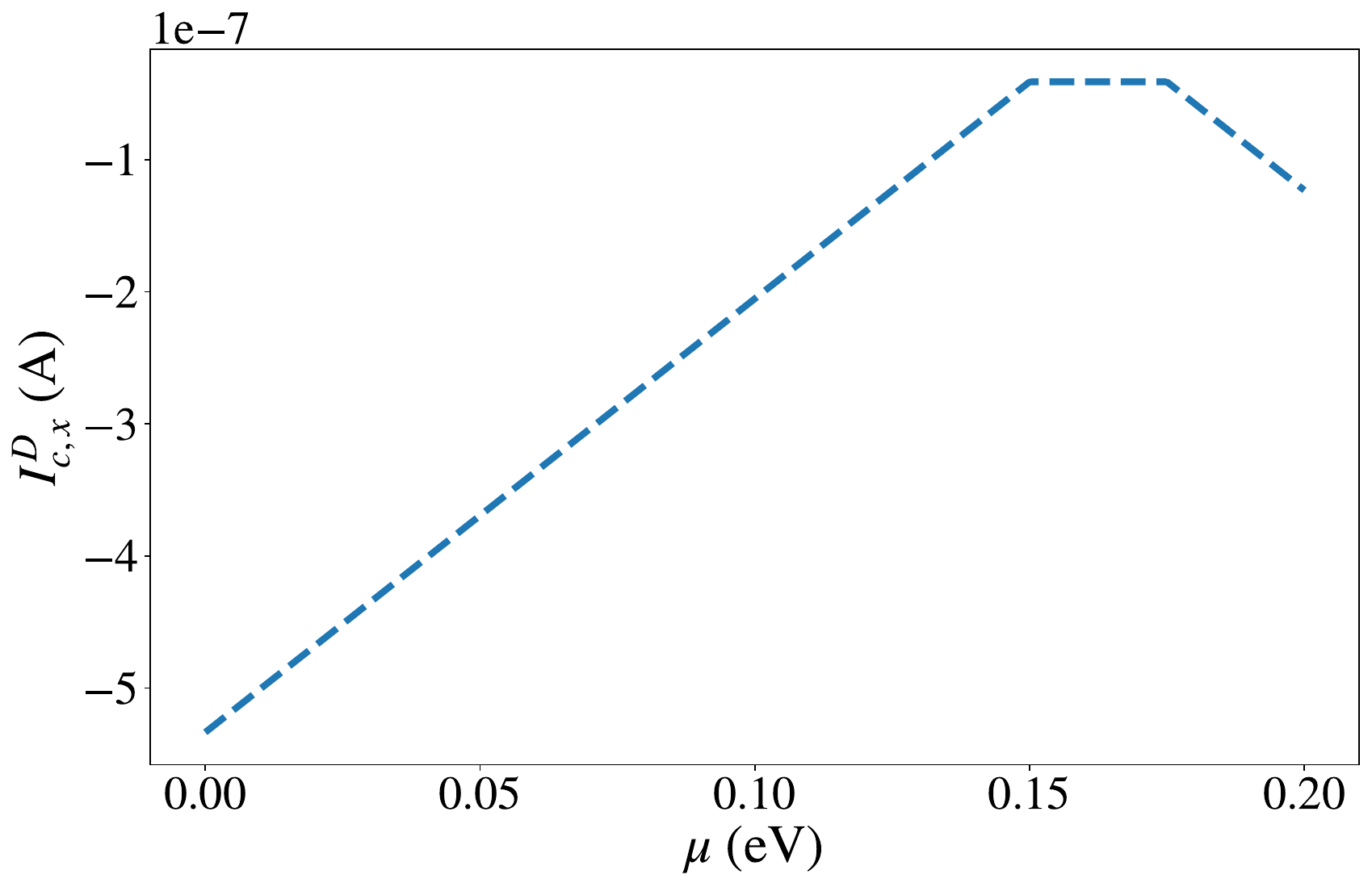}
\put(0,67){\rm{(a)}}
\end{overpic} & 
\begin{overpic}[width=0.33\linewidth]{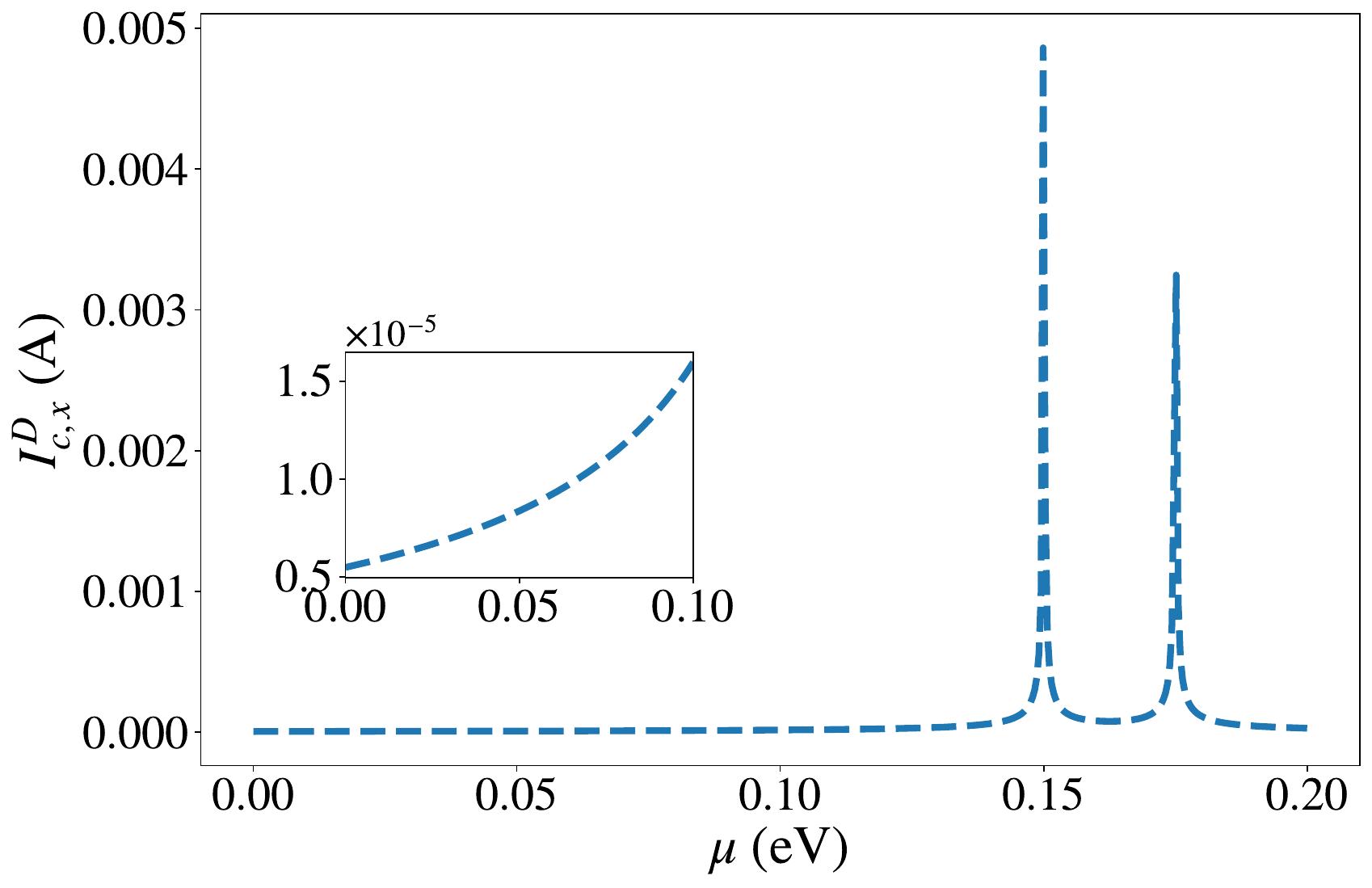}
\put(0,67){\rm{(b)}}
\end{overpic}& 
\begin{overpic}[width=0.33\linewidth]{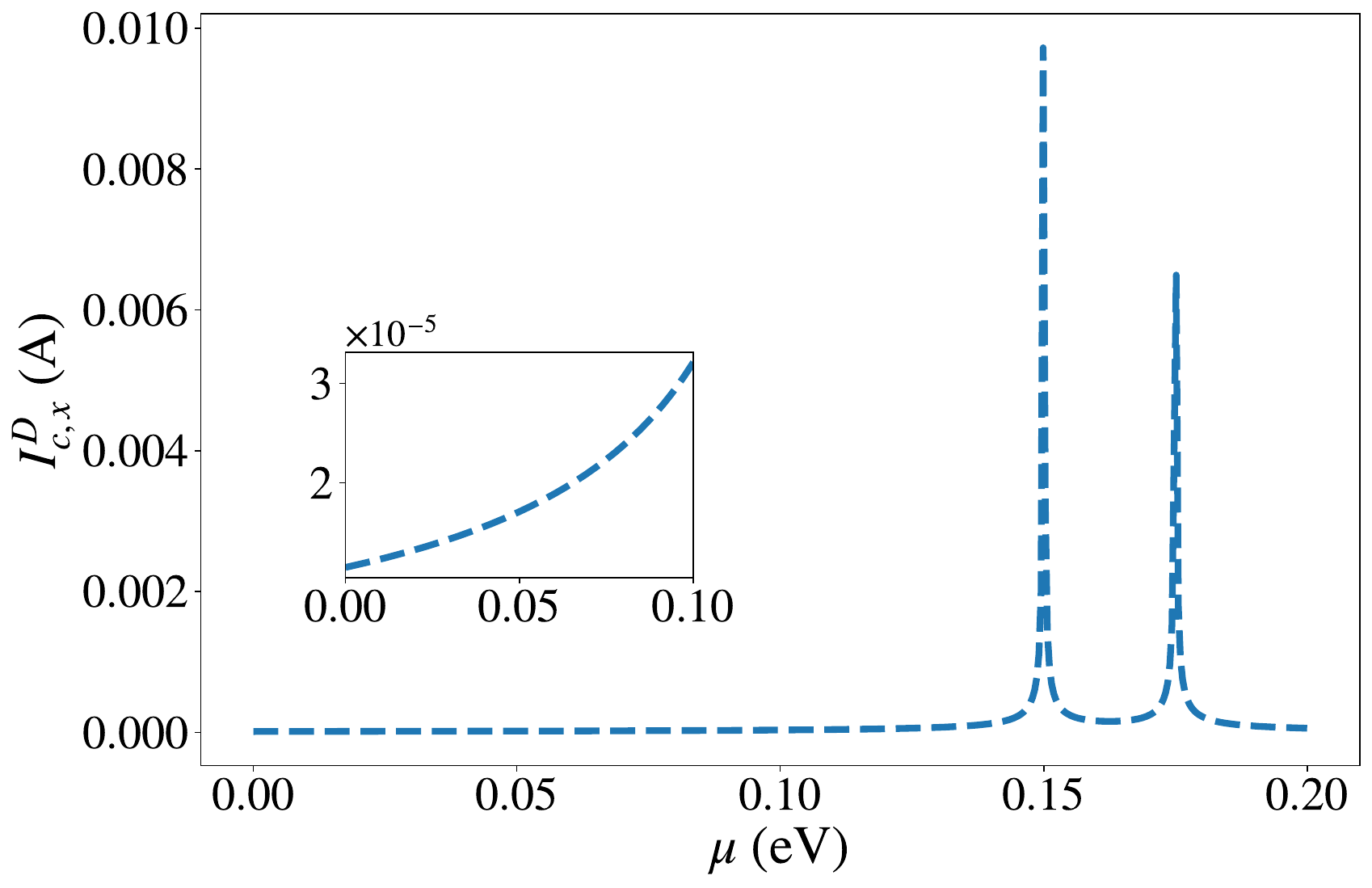}
\put(0,67){\rm{(c)}}
\end{overpic} 
\end{tabular}
\caption{Dependence of the Drude contribution to the charge current on the chemical potential. For the purpose of illustration, $\theta$ is held fixed at $\pi/3$;
(a) circular polarization ($\cos \phi = 0$),
(b) elliptical polarization ($\cos \phi = 1/2$),
and 
(c) linear polarization ($\cos \phi = 1$).}
\label{fig:ChemicalDC}
\end{figure}

\begin{figure}
    \centering
    \includegraphics[width=0.45\linewidth]{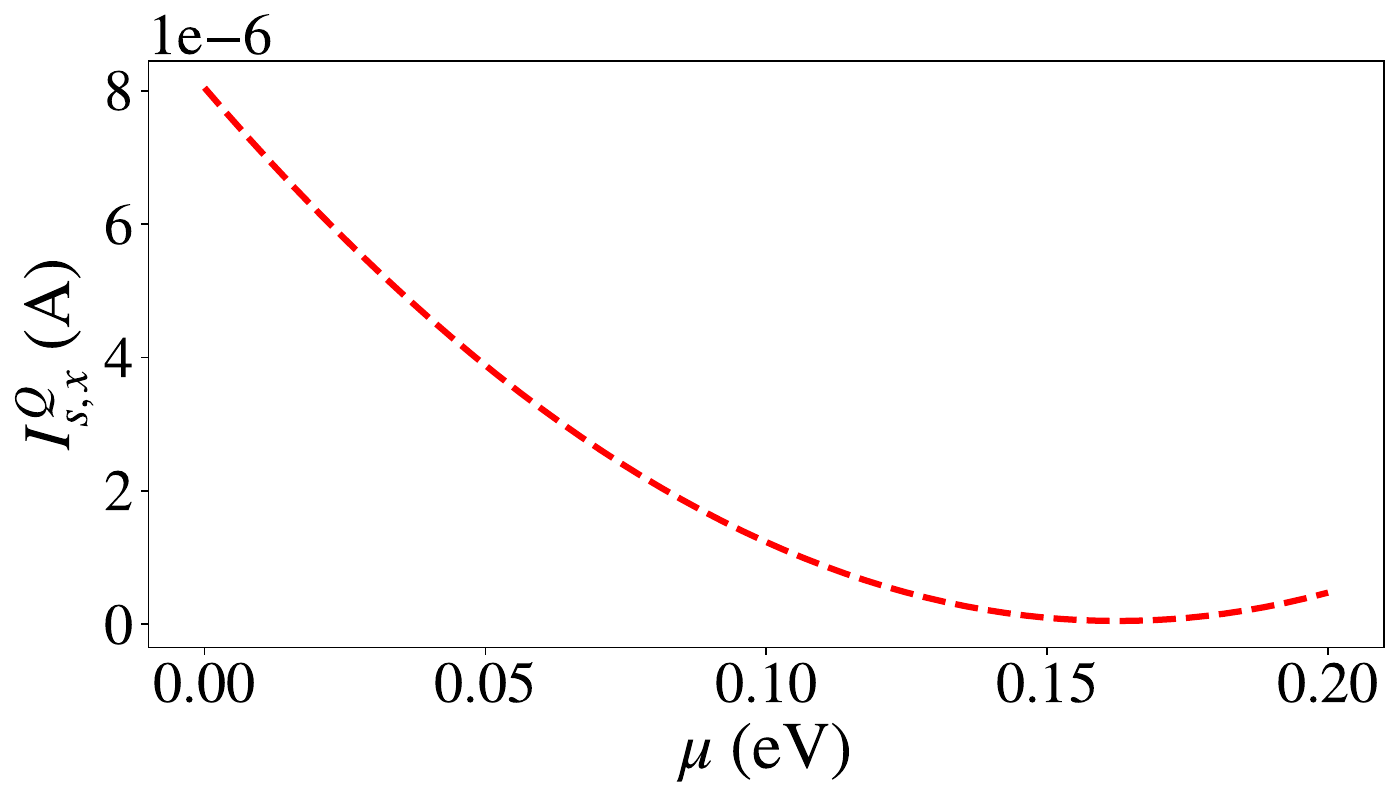}
    \caption{Dependence of the Berry curvature quadrupole contribution to the spin current on the chemical potential. }
    \label{fig:ChemicalSpinQ}
\end{figure}

Figures~\ref{fig:polarplotQ} and \ref{fig:polarplotD} show the charge current for three polarization choices. The polarization angle $\phi$ is held fixed, while $\theta$ is varied. 
The current contains two contributions, the Berry quadrupole and the Drude components, which are shown separately. Note that the induced spin current does not depend on the light polarization. Dependence on the chemical potential is illustrated in Figs.~\ref{fig:ChemicalQC}-\ref{fig:ChemicalSpinQ}.

\clearpage

\end{document}